\documentclass[pre,twocolumn,showpacs]{revtex4} 
\usepackage{t1enc,alltt,amssymb,amsmath,natbib}
\usepackage{graphicx}

\graphicspath{{images/}}

\usepackage{color}
\usepackage{amsmath}
\usepackage{subfigure}
\usepackage{overpic}
\begin{document}
\title{Spatio-temporal spectral analysis of a forced cylinder wake}
\author{Juan D'Adamo}\email{jdadamo@fi.uba.ar}
\affiliation{{Facultad de Ingenier\'ia Universidad de Buenos Aires (CONICET), Av. Paseo Col\'on 850, C1063ACV - Buenos Aires - Argentina}}
\author{Ramiro Godoy-Diana}\email{ramiro@pmmh.espci.fr}
\author{Jos\'e Eduardo Wesfreid}\email{wesfreid@pmmh.espci.fr}
\affiliation{{Physique et M\'ecanique des Milieux Het\'erog\`enes (PMMH), CNRS UMR 7636; ESPCI ParisTech; UPMC (Paris 6); Univ. Paris Diderot (Paris 7), 10 rue Vauquelin, 75231 Paris, Cedex 5, France}}


\begin{abstract}
The wake of a circular cylinder performing rotary oscillations is studied using hydrodynamic tunnel experiments at $Re=100$. Two-dimensional particle image velocimetry on the mid-plane perpendicular to the axis of cylinder is used to characterize the spatial development of the flow and its stability properties. The lock-in phenomenon that determines the boundaries between regions of the forcing parameter space were the wake is globally unstable or convectively unstable (see \cite{thiria2009} for a review) is scrutinized using the experimental data. A novel method based on the analysis of power density spectra of the flow allows us to give a detailed description of the forced wake, shedding light on the energy distribution in the different frequency components and in particular on a cascade-like mechanism evidenced for a high amplitude of the forcing oscillation. In addition, a calculation of the drag from the velocity field is performed, allowing us to relate the resulting force on the body to the wake properties.
\end{abstract}

\maketitle

\section{Introduction}
The simple geometry and the complex behavior of the flow around a circular cylinder at low Reynolds numbers ($Re=DU_0/\nu\lesssim 180$, where $D$ is the diameter of the cylinder, $U_0$ the free-stream velocity and  $\nu$ the kinematic viscosity of the fluid), makes it a prototypical 2D wake flow. The well-known B\'enard-von K\'arm\'an (BvK) vortex street \cite{benard1908,vonkarman1911} results from the destabilization of the steady flow in the wake of the cylinder, driven by the periodic shedding of opposite-signed vortices, that occurs above the threshold $Re_c\approx 47$ (see e.g. \cite{provansal1987,jackson1987}). Often used as a model for shear flow instabilities (see e.g. \cite{huerre1990,pier2002,chomaz2005,barkley2006,thiria_b_w_2007}), the cylinder wake gives a framework to study the distinctive features of spatially developing flows. In particular, for $Re>Re_c$ the velocity field in the whole flow domain oscillates with the same global frequency and its harmonics and, because the oscillation is spatially evolving, it can be characterized through the evolution of its envelope or \textit{global mode} of the instability as a function of the flow parameters \cite{wesfreid1996,zielinska1995}.\\
The cylinder wake has been also widely used to test methods of flow control using dynamic actuation, either in an open-loop sense \cite{protas2002} or with a feedback loop closed by the signal taken by a flow sensor \cite{siegel2006} and within an optimal control scheme \cite{protas_sty_2002}. A large subset of the existing literature about control strategies on the cylinder wake concerns the use of imposed oscillations, the case of in-line oscillations \cite{williamson1988} being intimately related to the problem of vortex-induced vibration (see extensive review by \cite{williamson2004}). For rotational oscillations, from the first visualizations of \cite{taneda1978} and the experimental work of \cite{tokumaru1991}, to the numerical \cite{cheng2001,shiels2001,choi2002} and experimental works \cite{thiria2006} that have followed, it has been shown in particular that the imposed rotational oscillation can significantly modify the geometry of the cylinder wake and hence the drag coefficient. The stability properties of the forced wake have been studied by \cite{thiria2007} using experimental measurements of the velocity field. They described in terms of the $(f,A)$ parameter space (the frequency $f$ and the amplitude $A$ of the oscillations) the two qualitatively different states that arise in the wake:  a spatial mode dominated by the BvK vortex street, meaning that the wake is globally unstable; and the so-called \textit{lock-in} regime, where the frequency is imposed by the forcing in the near wake, the amplitude of the oscillation rapidly decaying downstream with the characteristics of a convective instability.\\
The first goal of the present work is to refine the description of the transitions between the locked regime and the global instability, which have been described in previous works on this setup (see \cite{thiria2009} for a review), especially for forcing frequencies lower than the natural frequency. Particle image velocimetry (PIV) measurements allow us to study the spatial development of the forced flow, in particular the modification of the global mode represented by the velocity fluctuation envelope. Scaling laws in the forcing parameter space that universally describe the evolution of the global modes in the vicinity of the critical lines separating the lock-in and the globally unstable regions are derived from this data. A spectral analysis of the velocity fluctuations gives an alternative procedure to define the critical lines and confirms previous linear stability studies. The drag forces for each case are estimated from the PIV measurements using the so-called flux equation (see \cite{noca1997}). These results allow us to relate the resulting force on the body to its wake properties. 

\section{Experimental setup and parameters}

\begin{figure}
\begin{center}
\includegraphics[width=0.7\linewidth]{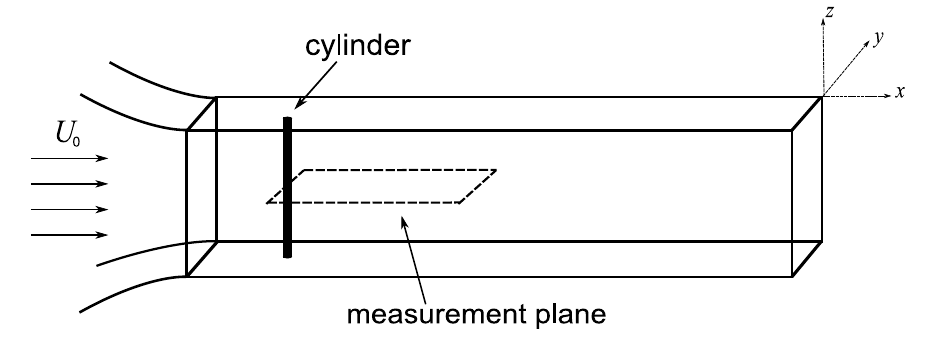}
\includegraphics[width=0.28\linewidth]{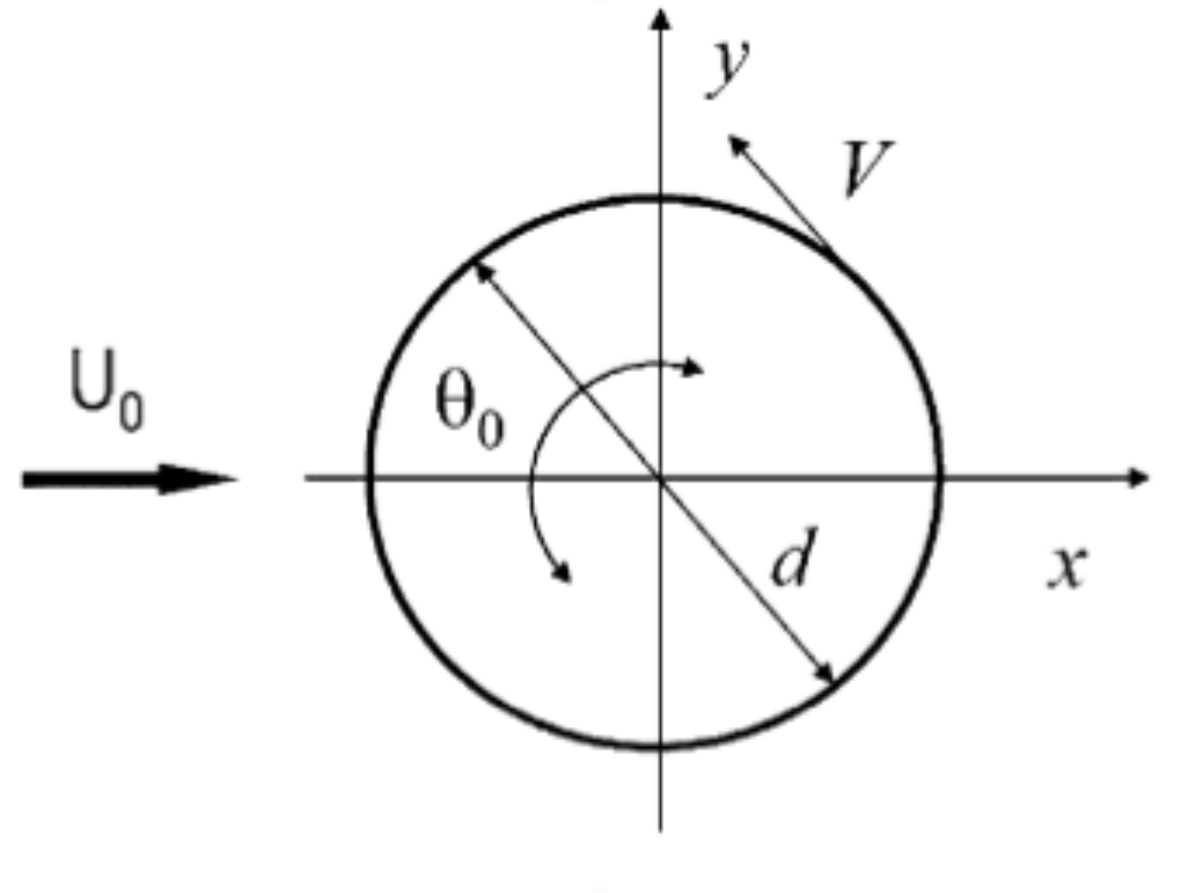}
\includegraphics[width=0.7\linewidth,trim= 10mm 28mm 15mm 35mm,clip]{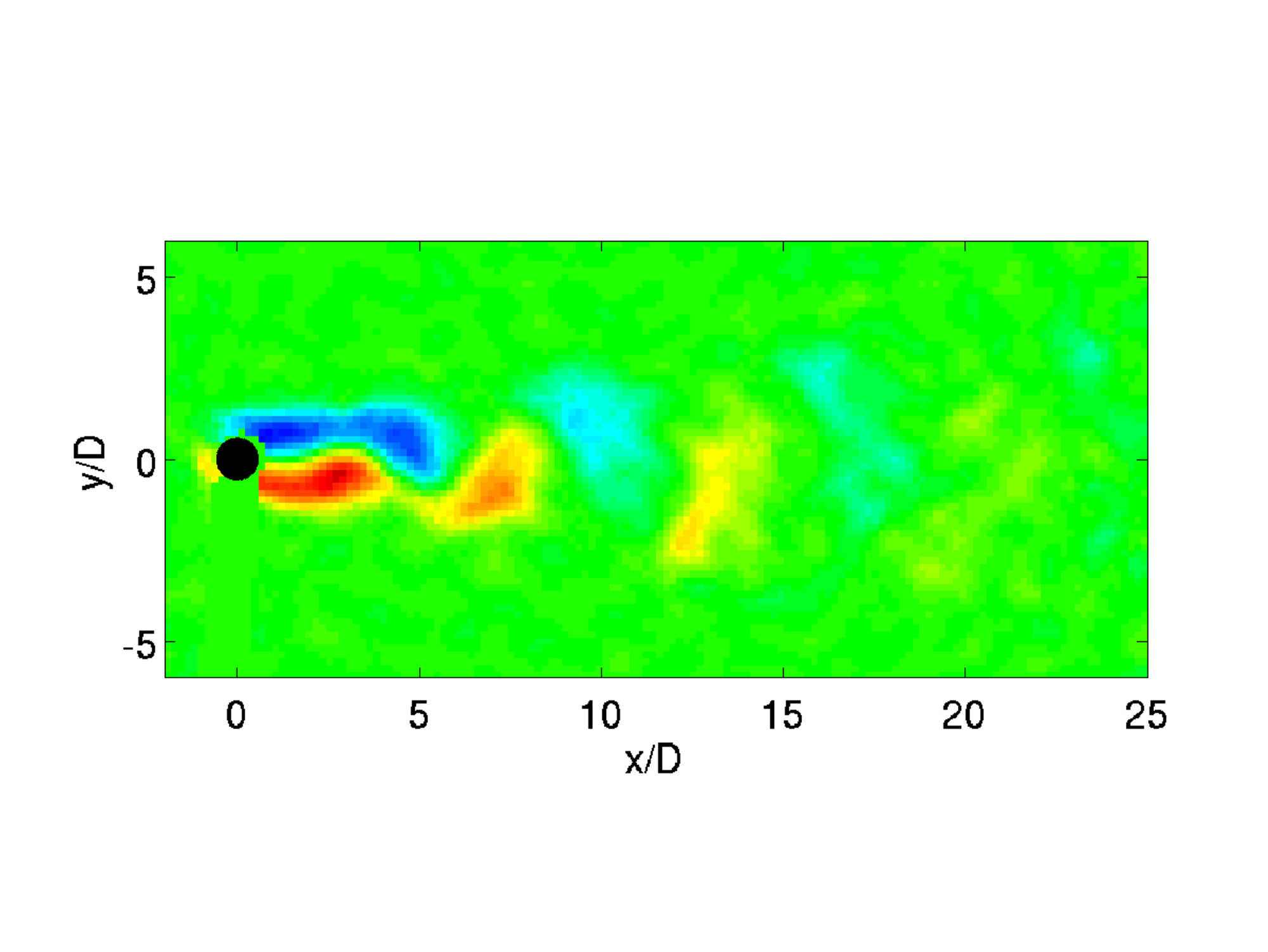}\vspace{-5mm} 
\end{center}
\caption{(Color) Top: Experimental setup and definition of the forcing parameters. Bottom: Vorticity field calculated from PIV measurements for the non-forced case showing the B\'enard-von K\'arm\'an vortex street.}
\label{fig_exp_setup}
\end{figure}

The experimental setup shown in Fig.~\ref{fig_exp_setup} is the same used by \cite{thiria2006}. A circular cylinder of diameter $D=5mm$ and span of $\approx 20D$ is placed in a low-speed hydrodynamic tunnel with a $100\times 100\mathrm{mm}$ cross-section. The cylinder span thus practically covers the whole height of the tunnel and the ratio of the cylinder diameter to the tunnel section width is of $1/20$. The cylinder can perform rotational oscillations driven by a stepper-motor placed below the platform of the test section on a submerged 'technical section' of the tunnel, imposing a controlled forcing on the flow. The Reynolds number is set to 100, defining it using the up-stream velocity in the center of the tunnel as  $U_0$. The measured natural vortex shedding frequency was $f_0=0.63$ Hz, so the Strouhal number $St= f_0 D / U_0$ is approximately 0.15. 
Quantitative measurements were performed using 2D particle image velocimetry (PIV) on a horizontal plane placed at mid-span of the cylinder (see the vorticity field for the non-forced case which corresponds to the well known B\'enard-von K\'arm\'an (BvK) vortex street in Fig.~\ref{fig_exp_setup}). Image acquisition and PIV calculation were done using a LaVision$\textsuperscript{\textregistered}$ system composed of an ImagerPro $1600 \times 1200$ CCD camera with a 12-bit dynamic range recording double-frame images at $11$Hz  and a two rod Nd:YAG (15mJ) pulsed laser synchronized by a customized PC using LaVision DaVis 7.1 software. Laser sheet width was about 1 mm in the whole $100\times 80$ mm imaging region. The time lapse between the two frames of each image pair used for PIV was set to $\Delta t=12$ms and sets of 500 snapshots give a frequency resolution of  $\Delta f =0.02$Hz.

\begin{figure}
\includegraphics[width=1\linewidth,trim= 11mm 1mm 15mm 5mm,clip]{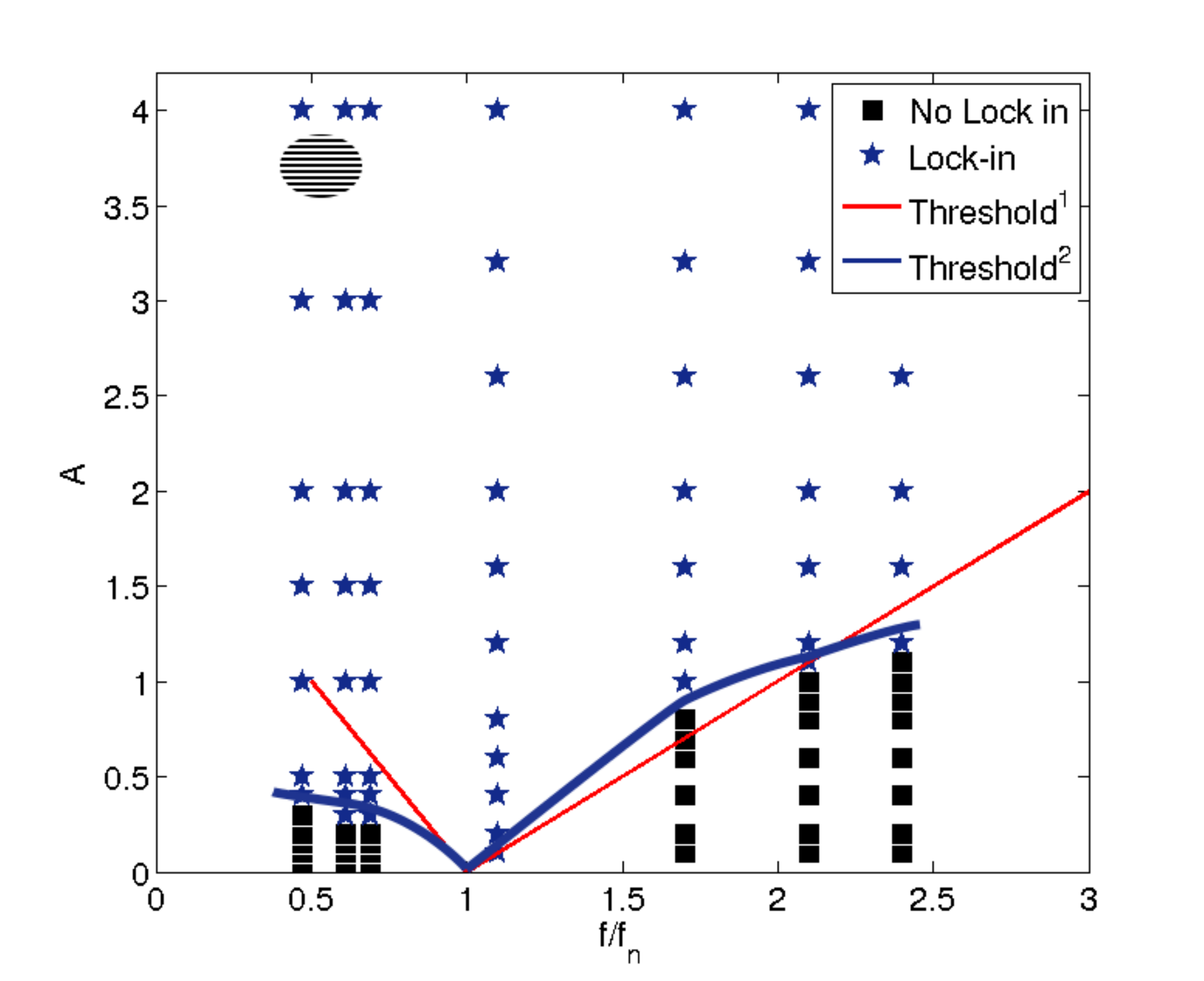}
\caption{(Color) Forcing parameter space. The points represent experiments where: $\blacksquare$ no lock-in is observed; {\color{blue}$\bigstar$} lock-in is observed. The blue line is the lock-in threshold estimated from the present experiments. The red line stands for the observations of  \cite{thiria2007}. Striped region indicates a maximum drag reported by \cite{bergmann2005}.}
\label{totalcasos}
\end{figure}

The rotational oscillation of the cylinder is prescribed by a forcing function of frequency $f$ and amplitude $\theta_0$ that can be written as $\theta(t)=\theta_0\cos(2\pi f t)$, which allows the forcing to be unequivocally described using two independent non-dimensional parameters as did by \cite{taneda1978}: the forcing amplitude $A=u_{\theta\mathrm{max}}/U_0$, where $u_{\theta\mathrm{max}}=D\pi f\theta_0$ is the maximal azimuthal velocity of the rotational oscillation; and the ratio $f/f_0$. We explore 100 forcing cases and they are represented in Fig.~\ref{totalcasos}. We define $A_c$ as the value of the amplitude for the transition between lock-in and  non-locked state (critical lines). These values are obtained from spectral analysis developed on section \ref{spec_study} and Fig~\ref{totalcasos} compares them with previous works based on linear stability from \cite{thiria2007} at $Re=150$. In order to compare the lock-in region with drag force estimation (section \ref{drag_estim}), we include in the figure the maximum drag found in \cite{bergmann2005} numerical work for $Re=200$  

\section{Revisiting vortex patterns and global modes in the wake}

\begin{figure}
     \begin{tabular}{l r} 
	\hspace{0.06\linewidth} $f_f=0.69f_0$ & \hspace{0.25\linewidth}$f_f=2.40f_0$
     \end{tabular}
\begin{overpic}[width=0.49\linewidth,trim= 10mm 43mm 15mm 35mm,clip,unit=1mm]%
{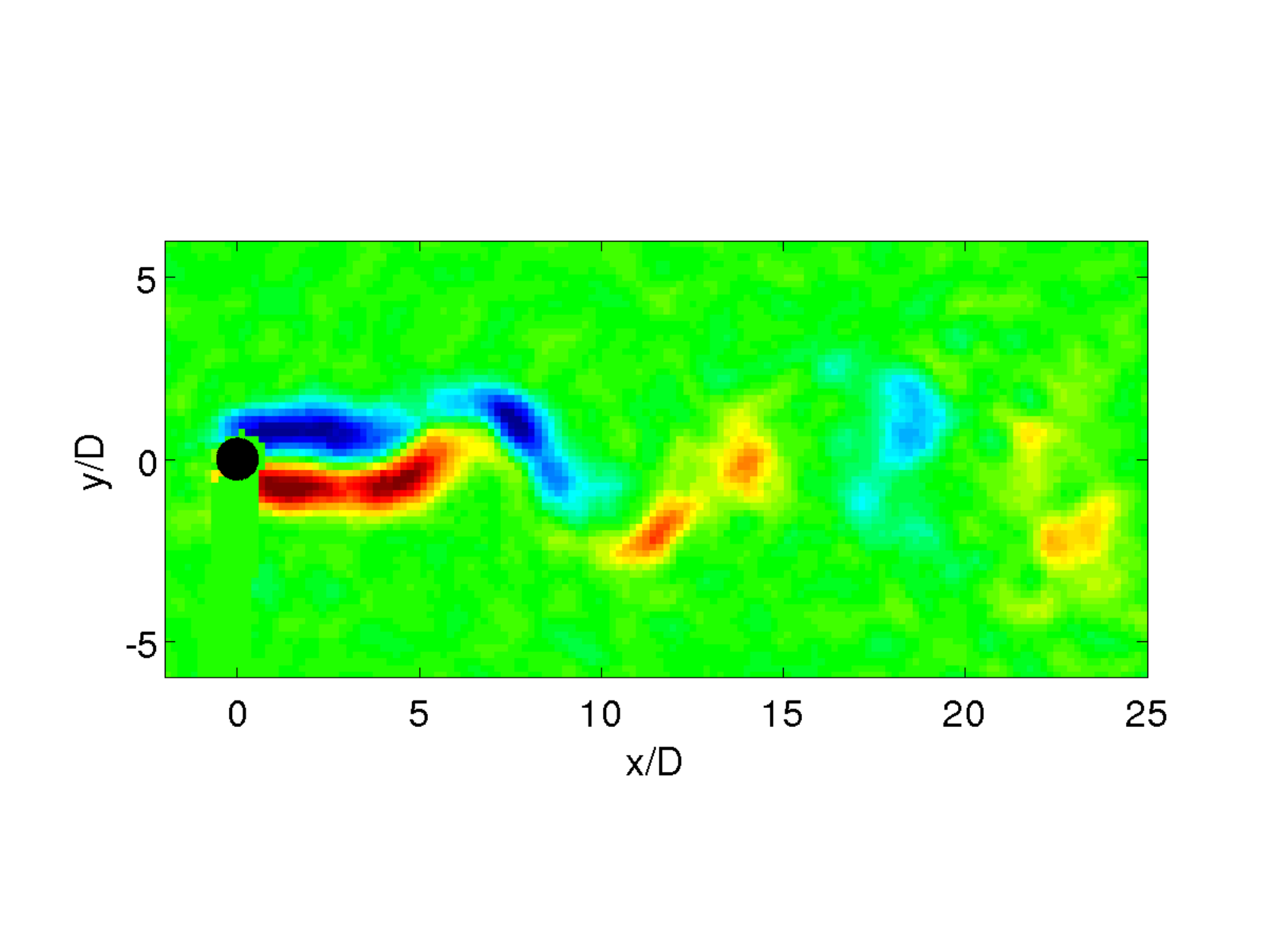}
\put(1,30){\footnotesize (a)}
\end{overpic}
\begin{overpic}[width=0.49\linewidth,trim= 10mm 43mm 15mm 35mm,clip,unit=1mm]%
{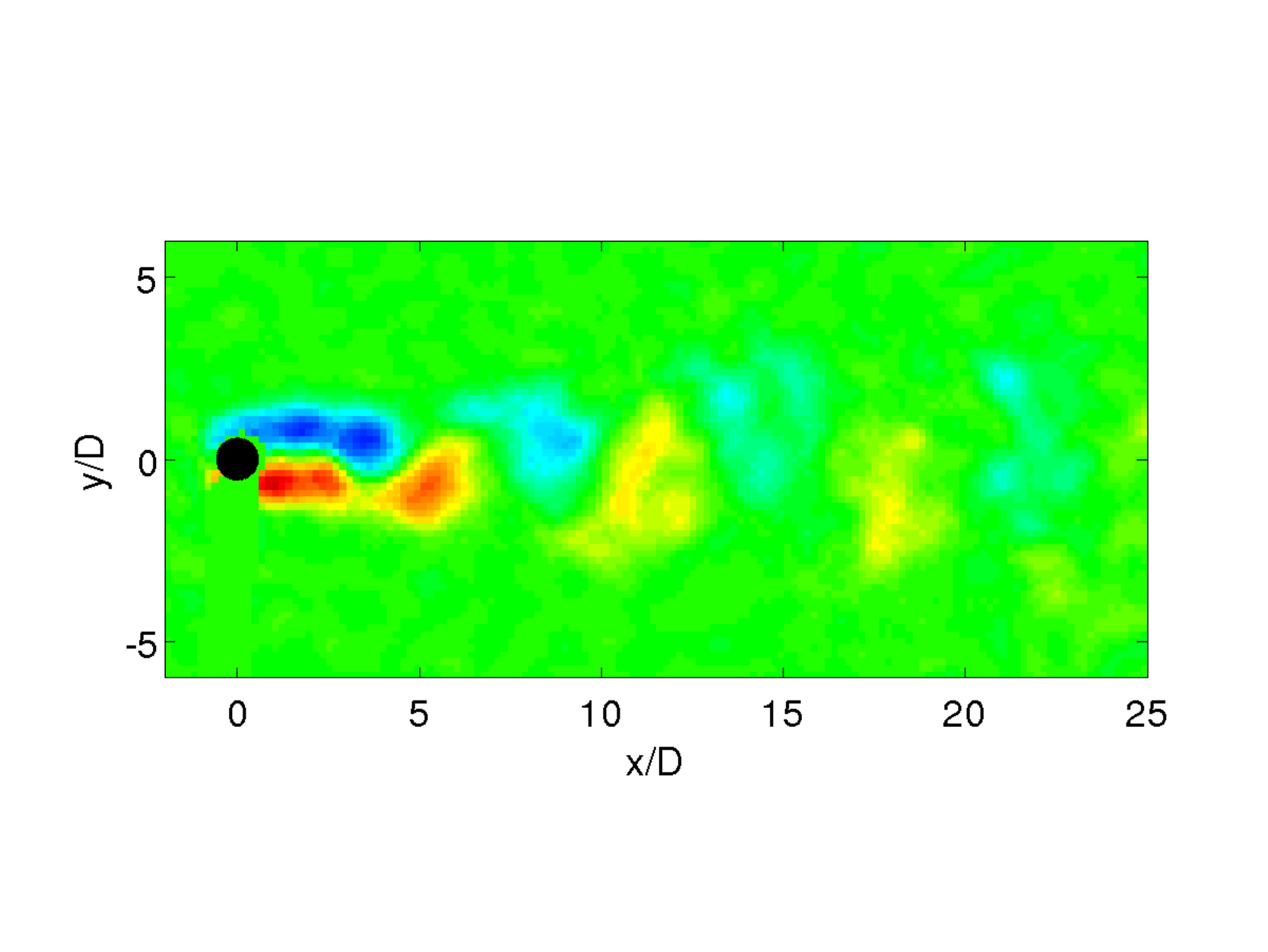}
\put(1,30){\footnotesize (e)}
\end{overpic}
\begin{overpic}[width=0.49\linewidth,trim= 10mm 43mm 15mm 35mm,clip,unit=1mm]%
{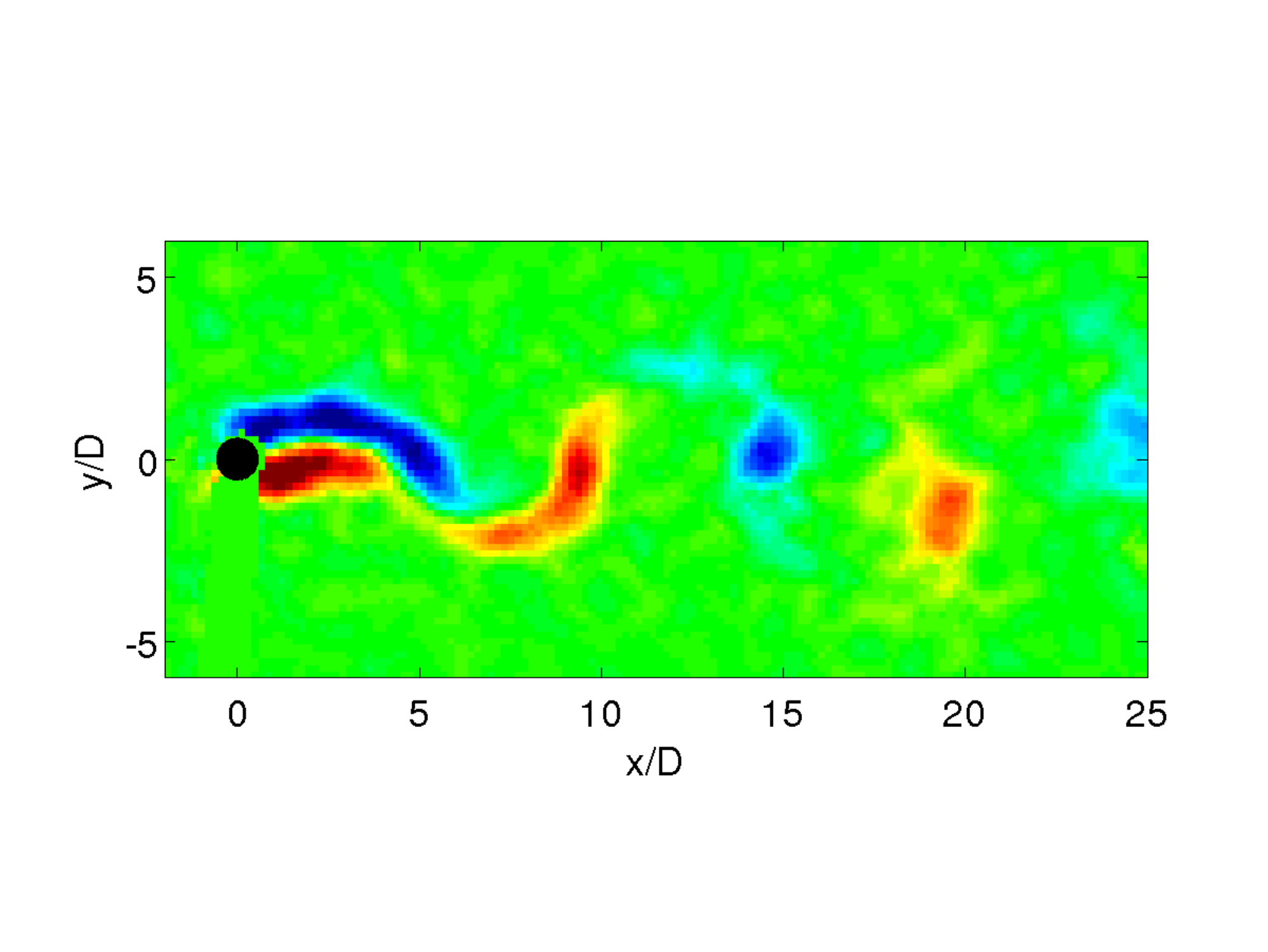}
\put(1,30){\footnotesize (b)}
\end{overpic}
\begin{overpic}[width=0.49\linewidth,trim= 10mm 43mm 15mm 35mm,clip,unit=1mm]%
{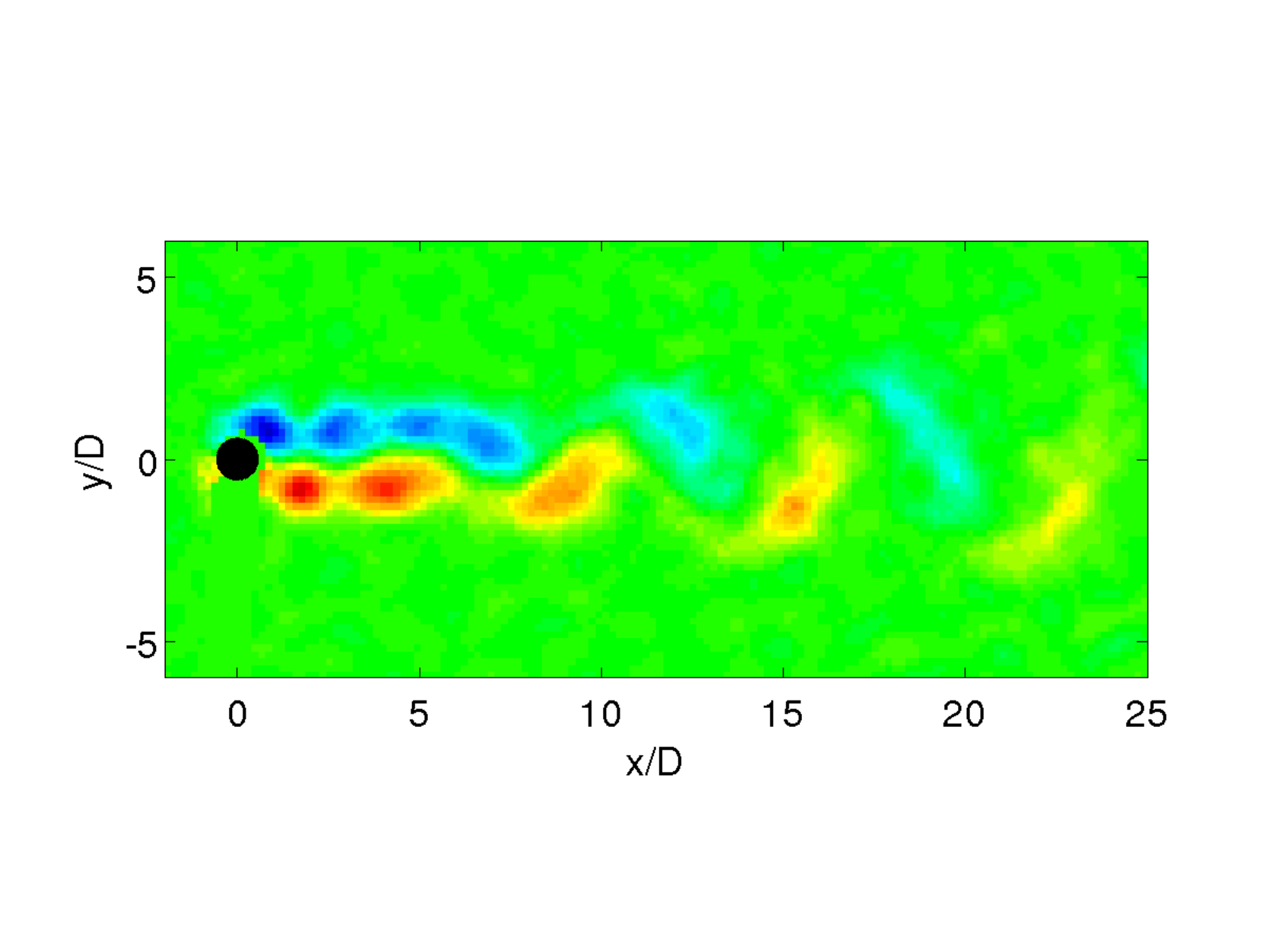}
\put(1,30){\footnotesize (f)}
\end{overpic}
\begin{overpic}[width=0.49\linewidth,trim= 10mm 43mm 15mm 35mm,clip,unit=1mm]%
{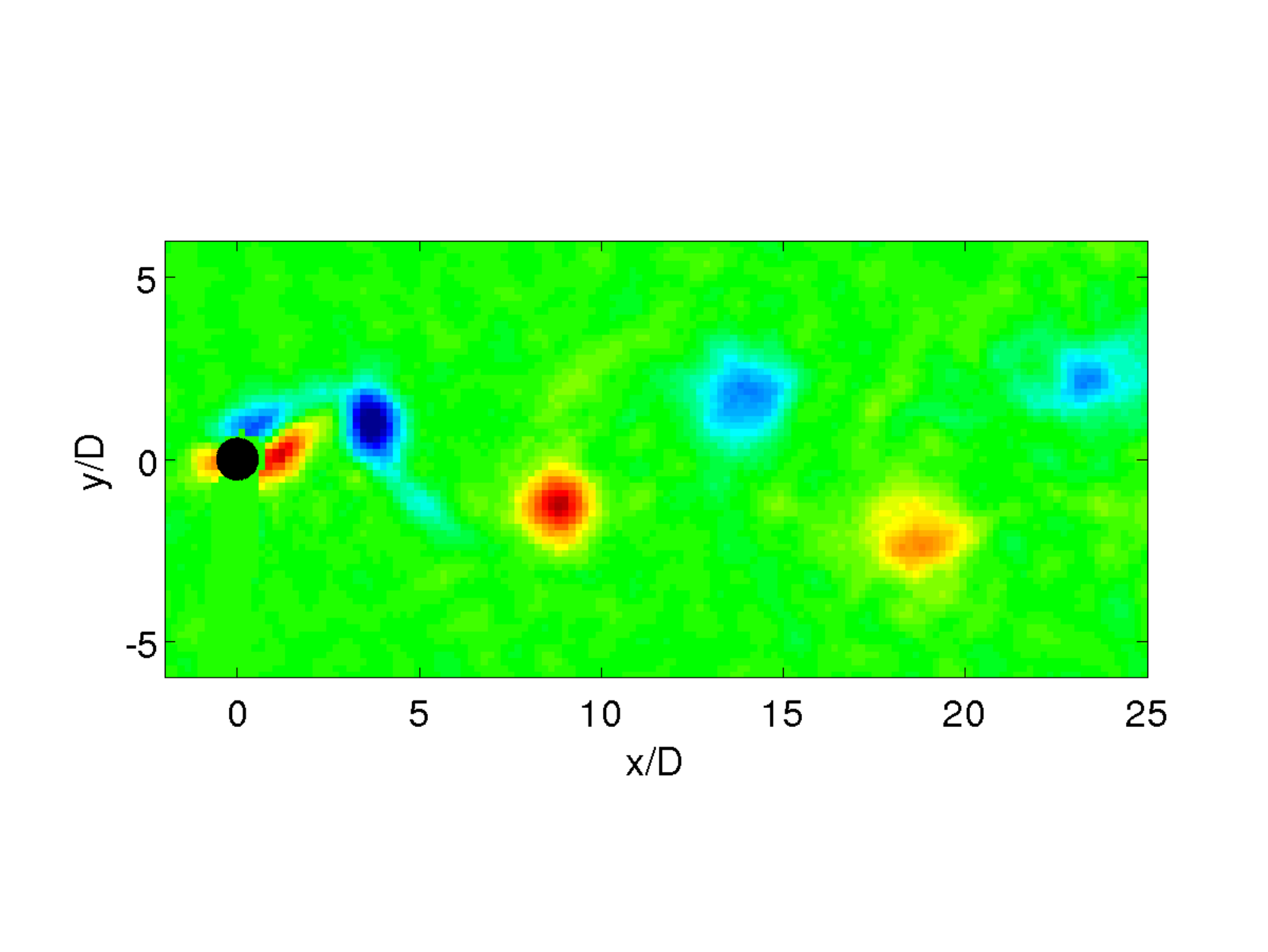}
\put(1,30){\footnotesize (c)}
\end{overpic}
\begin{overpic}[width=0.49\linewidth,trim= 10mm 43mm 15mm 35mm,clip,unit=1mm]%
{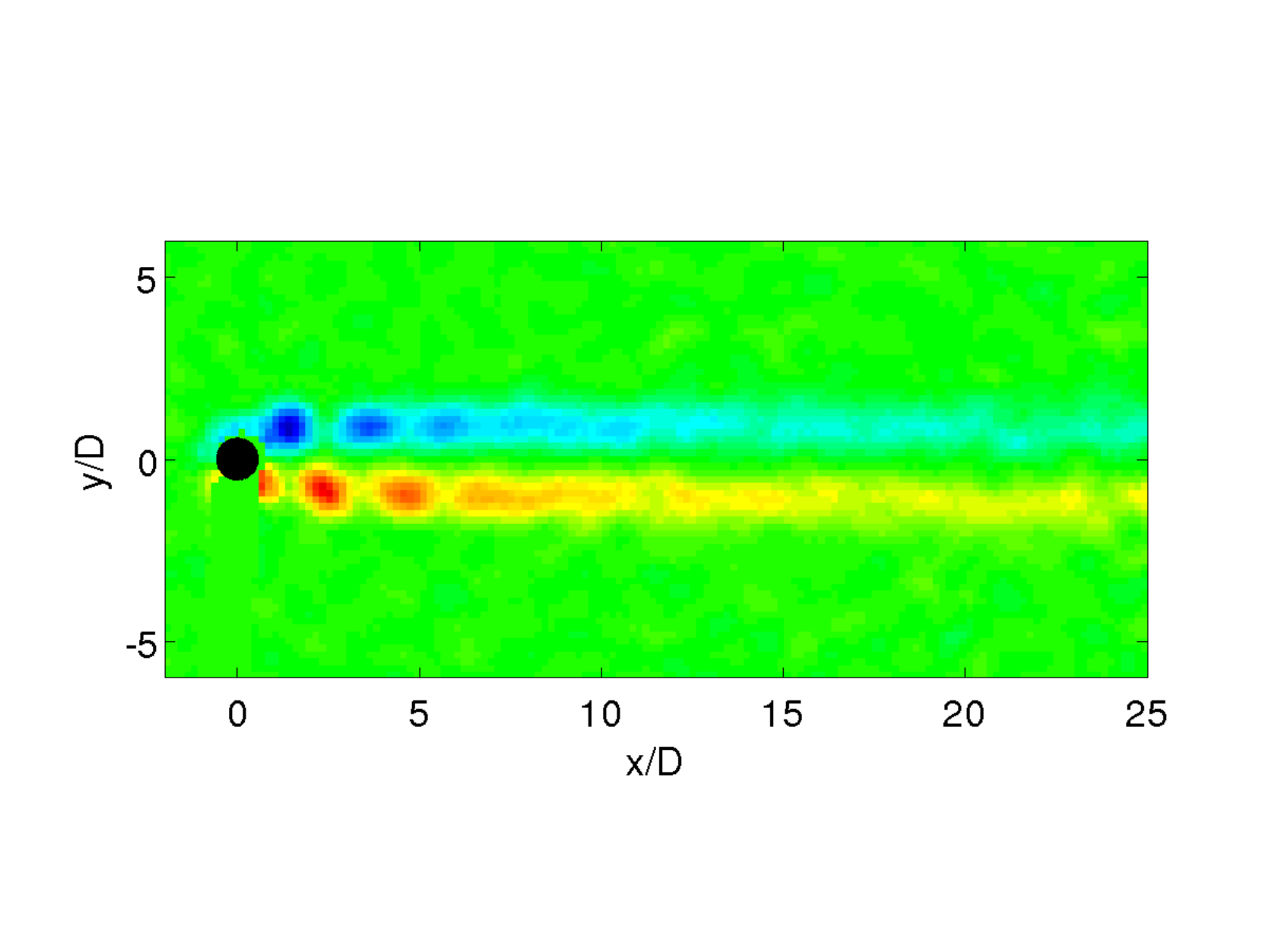}
\put(1,30){\footnotesize (g)}
\end{overpic}
\begin{overpic}[width=0.49\linewidth,trim= 10mm 43mm 15mm 35mm,clip,unit=1mm]%
{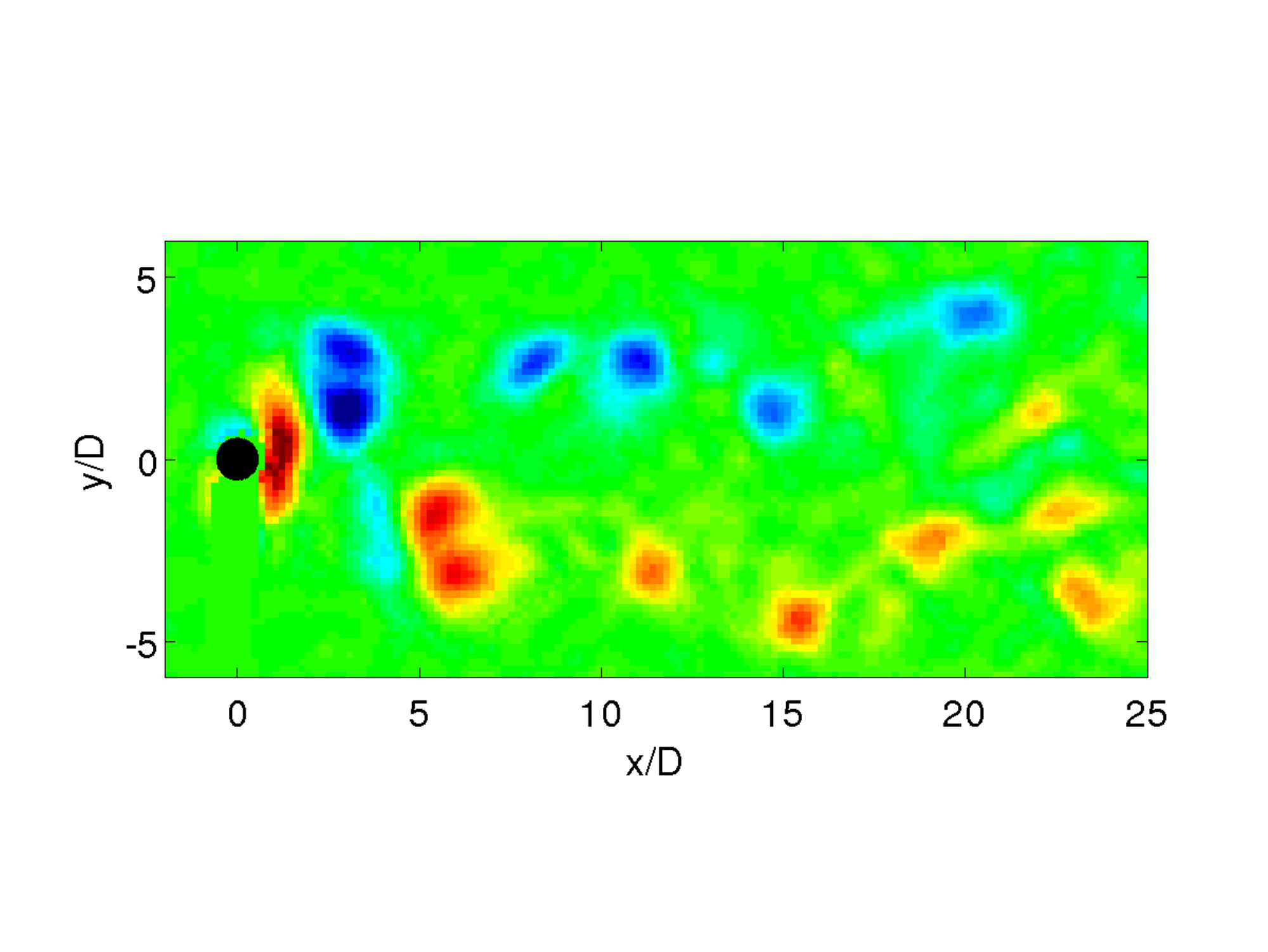}
\put(1,30){\footnotesize (d)}
\end{overpic}
\begin{overpic}[width=0.49\linewidth,trim= 10mm 43mm 15mm 35mm,clip,unit=1mm]%
{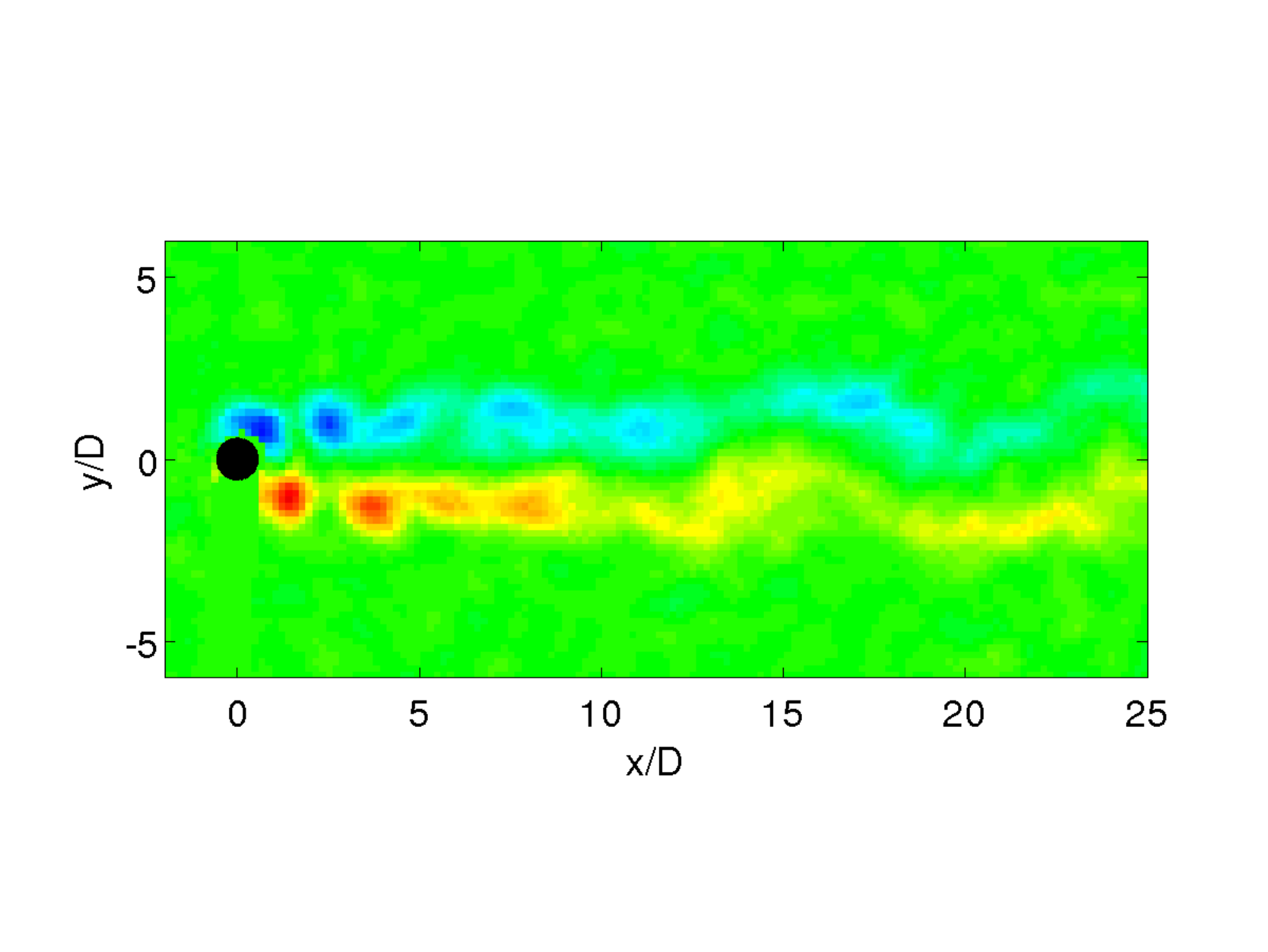}
\put(1,30){\footnotesize (h)}
\end{overpic}
   \caption{(Color) Vortex structures from vorticity contours for $f_f=0.69f_0$
(\textit{Left}) and $f_f=2.40f_0$ (\textit{Right}): (a) $A=0.10<A_c$, (b)$A=0.40>A_c$, (c) $A=1.00>A_c$, (d) $A=3.2$,
(e) $A=0.40<A_c$, (f) $A=0.80<A_c$, (g)  $A=1.60>A_c$, (h) $A=4.00$. [See text]}\label{vortex_str}
\end{figure}

A first step to characterize a regime that arises under forcing is by identifying its vortex patterns, as initially done in the early works of \cite{williamson1988}. In Fig.~\ref{vortex_str}, we show the wake patterns for two different forcing frequencies and various amplitudes. In the left column of Fig.~\ref{vortex_str}, $f_f=0.69f_0$ and $A_c\sim 0.3$. We observe at low amplitude (a) $A=0.10$ that the flow is not entirely synchronized with the natural frequency but the formation region is modulated by the forcing frequency.  Increasing the amplitude  to (b) $A=0.40$, the flow is locked on the forcing frequency and vortices are shed in a shorter distance. The vortex structure remains the same as the non-forced case as each half a cycle a vortex is fed into the wake. For $A=1.00$ the convective character of the instability is clearly evidenced by  the intense vortices shed in the vicinity. For $A>1$ two vortex rows form as the vortex cores move away from the cylinder. Once this pattern is attained, under further augmentation of the amplitude forcing, the vortex structures start splitting in a cascade-like pattern as can be seen in Fig.~\ref{vortex_str} (d). The spectral analysis performed below gives more insight on this phenomenon. The right column of Fig.~\ref{vortex_str} corresponds to the case of $f_f=2.40 f_0$ and $A_c\sim 1$, and the evolution of the wake pattern under forcing develops in a different way. For  $A=0.40$ (Fig.~\ref{vortex_str}(e)), the wake aspect does not differ from natural BvK vortex shedding except that in the near wake the forcing wavelength slightly modulates the vortex formation. As the forcing amplitude increases the shedding distance from the cylinder also grows: in Fig.~\ref{vortex_str}(f) it is $\sim 10D$, for $A=0.80<A_c$, i.e. approximately twice of what is observed at $A=0.40$. Indeed, vortices  of the same sign coalesce in the near wake constituting a long formation region until they are shed in a BvK-like pattern farther downstream. Once the lock-in threshold amplitude ($A_c\sim 1$) is exceeded, the coalescence pattern prevails and two vortex sheets are formed stabilizing the wake (see Fig.~\ref{vortex_str}(g)). Higher forcing amplitudes, for instance $A=4$ in Fig.~\ref{vortex_str}(h),  destabilize again the wake and a BvK-like pattern reappears in the far wake.

\subsection{Global mode shape}

\begin{figure}
\centering
\begin{overpic}[width=.70\linewidth,unit=1mm]%
{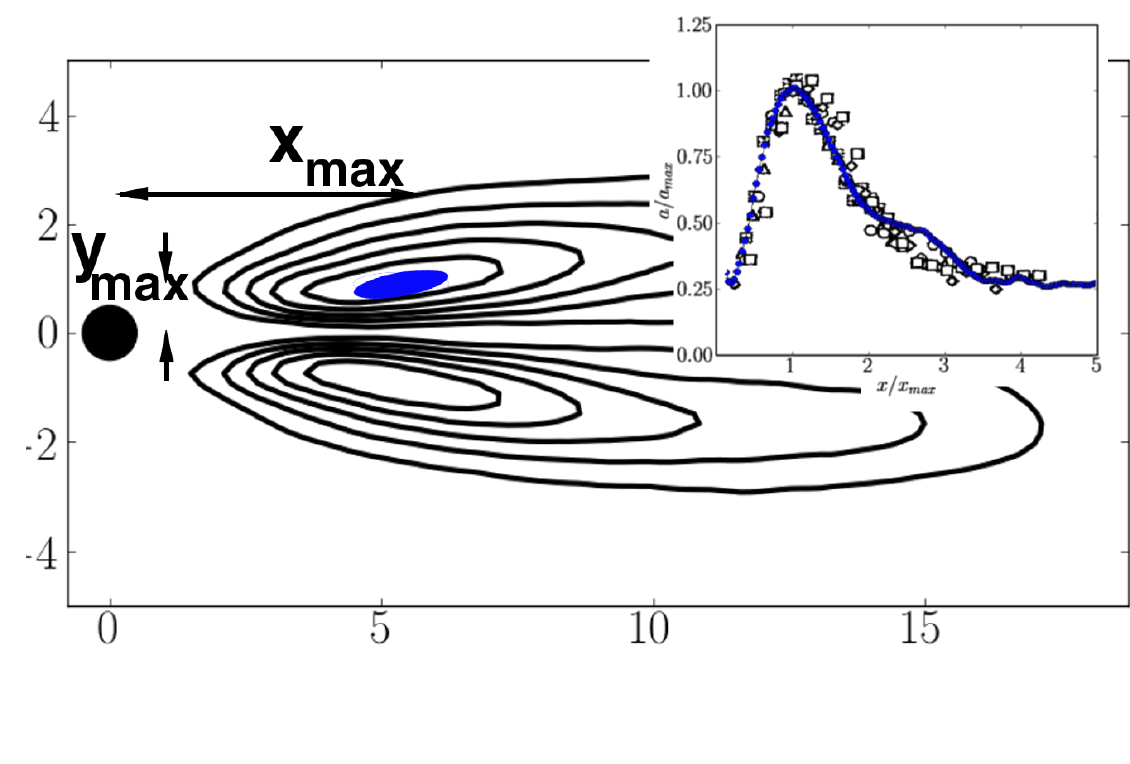}
\put(-10,75){\footnotesize (a)}
\end{overpic}\\
\hspace{-7mm}\begin{overpic}[width=.84\linewidth,unit=1mm]%
{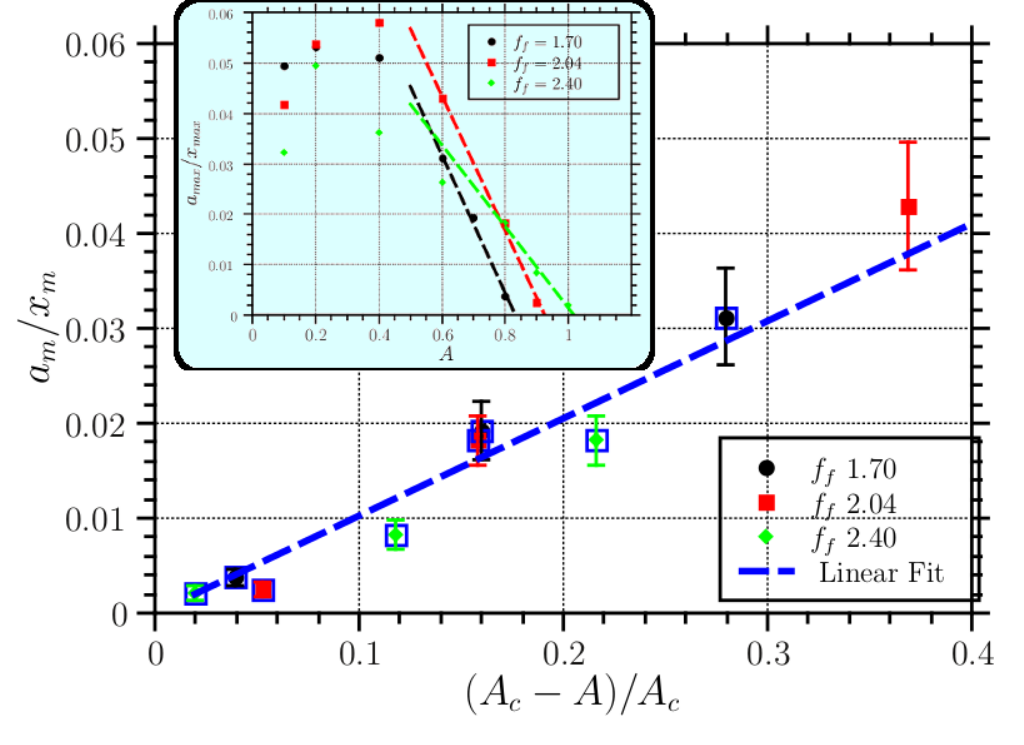}
\put(1,60){\footnotesize (b)}
\end{overpic}\\
\begin{overpic}[width=.78\linewidth,unit=1mm,trim=10mm 9mm 19mm 9mm,clip]%
{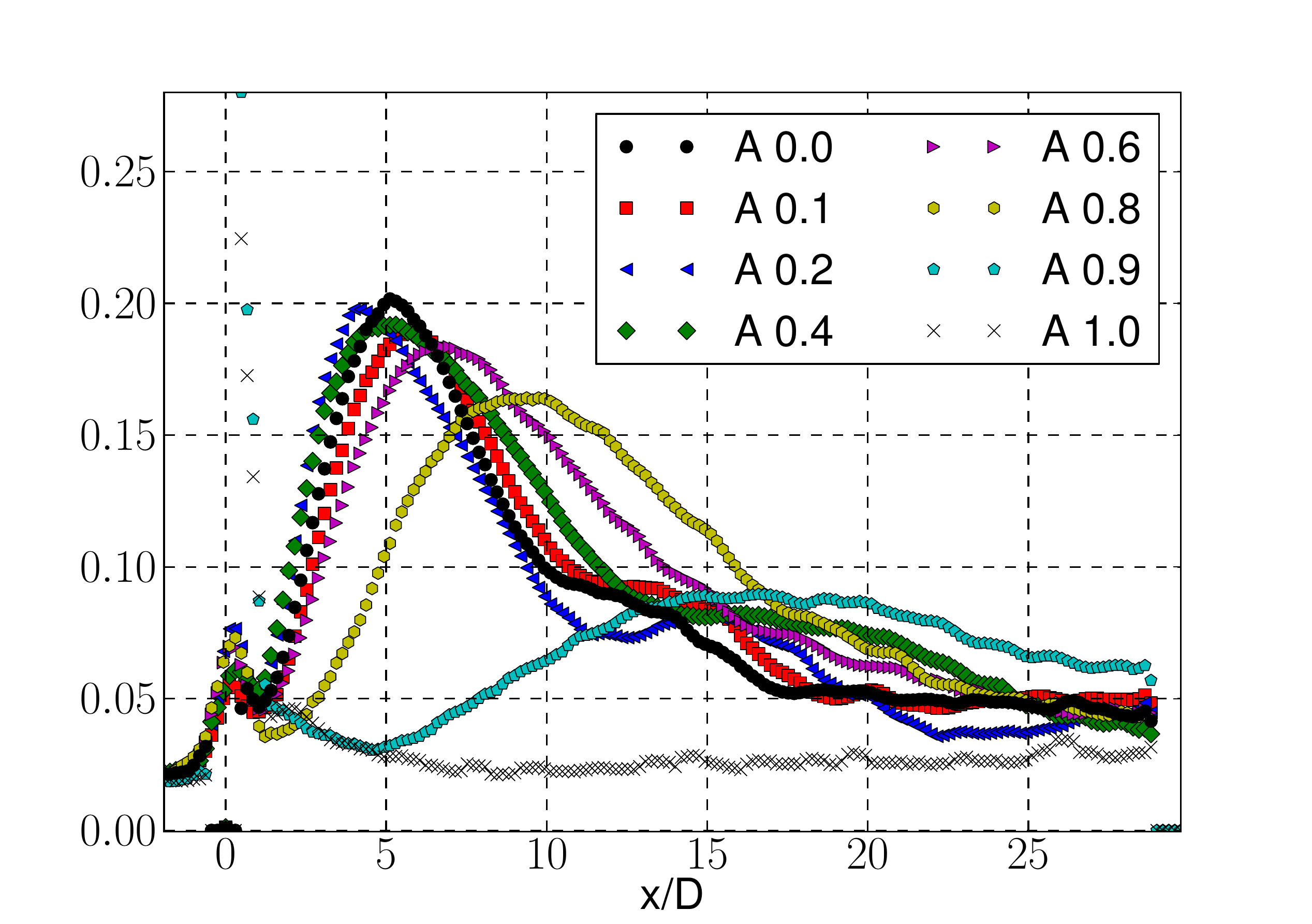}
\put(-5,60){\footnotesize (c)}
\end{overpic}
\begin{overpic}[width=.78\linewidth,unit=1mm,trim=10mm 0mm 19mm 9mm,clip]%
{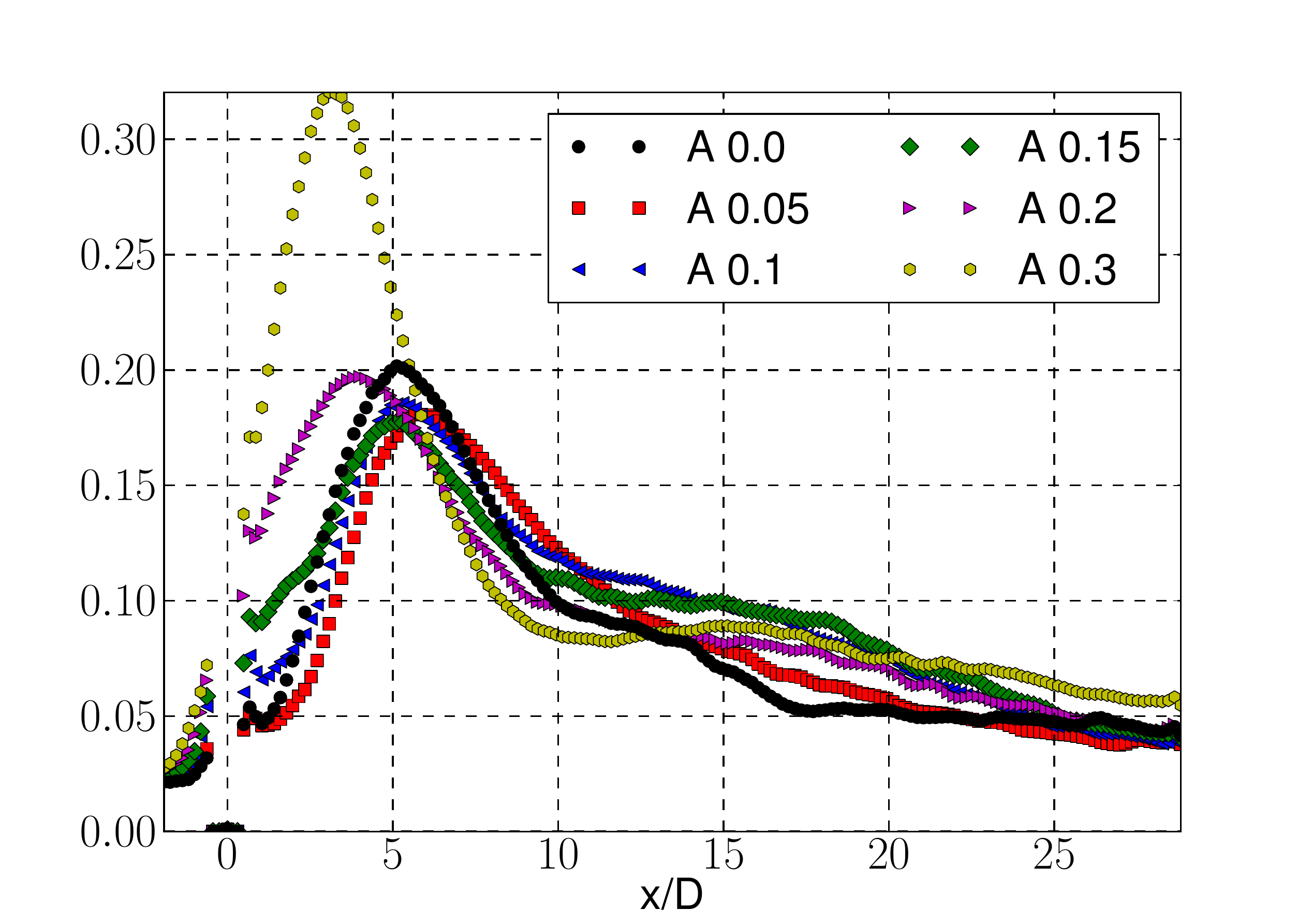}
\put(-5,60){\footnotesize (d)}
\end{overpic}
\caption{(Color) (a) Global mode spatial structure (from \cite{wesfreid1996}): isocontours of $<u_{x}>$ for the non-forced flow around a cylinder and shape of the global mode envelope evaluated at the maximum (blue region) $y=y_{max}$ in the inset (the solid line is from the present experiments and the points form \cite{wesfreid1996}). From the present experiments: envelope of the global mode given by the amplitude of the $u_x$ fluctuations, as a function of the downstream distance $x$ to the cylinder and for different values of the forcing amplitudes $A$ for $f_f=2.40f_0$ (c) and $f_f=0.69f_0$ (d); and (b) scaling laws for the spatial envelopes for $f_f>f_o$. Estimation of $A_c$ from $a_{max}/ x_{max}$ versus $A$ in the inset and scaled result for $a_{max}/x_{max}$ in (b).}
\label{global_modes}
\end{figure}

Wake flows can be analyzed as a propagating wave with an amplitude (determined by the fluctuating component of velocity) that grows from the origin, reaches a maximum and decays afterwards. The spatial envelope of this coherent oscillation gives the amplitude of the so-called global mode, for which the dominant contribution is given by the first harmonics. Previous works \cite{wesfreid1996,zielinska1995} studied scaling laws for the global mode in wake flows near the threshold $Re_c$. A typical contour is presented on Fig.~\ref{global_modes}(a) where its maximum amplitude, $a_{max}$ at $(x_{max}, y_{max})$ coordinates, is highlighted as it represents an important parameter for scaling . A synthesis for the global mode properties is represented by its envelope, Fig.~\ref{global_modes}(a, inset), that corresponds to the position of $x_{max}$ the maximum amplitude of the mode at $y=y_{max}$. The authors proposed that both $a_{max}$ and $x_{max}$ follow scaling laws such as $a_{max}\sim \varepsilon^\beta$ and $x_{max}\sim \varepsilon^\nu$ in the vicinity of the bifurcation. As $\varepsilon$ represents a control parameter that measures the distance to the threshold they determined that $a_{max}\simeq    (Re - Re_c )^{1/2}$ and $x_{max}\simeq (Re - Re_c )^{-1/2 }$.

The existence of global modes is a consequence of the existence of an absolute instability region, which is modified by  the forcing. The region disappears after lock-in is attained. For $f>f_0$ this is discussed in \cite{thiria2007} and revisited in the present experiment. Fig.~\ref{global_modes}(c) shows the vanishing of the global mode as its envelope broadens and its maximum diminishes. Once the lock-in threshold amplitude is exceeded, the mean energy  of the fluctuations $u_{rms}=\int_0^T u_x'^2 dt /T$ is driven exclusively by the forcing, the curve strongly decreases from the cylinder to the wake. This observed envelope is expected to decay exponentially with the form $A e^{x/\xi}$, where $\xi$ has the same critical behavior as $x_{max}$ but with $|\epsilon|$, as it was studied in early works by \cite{wesfreid1979}.\\
 The picture is less clear for the cases where $f_f<f_0$, shown in Fig.~\ref{global_modes}(d). While small forcing amplitudes determine the global mode maximum to decrease, for $A>0.15$ the envelope increases, and it does it significantly after the lock-in threshold. 
 
\subsection{Scaling laws}
The evolutions of the maximum of the global mode envelope $a_{max}$ and its position $x_{max}$ can be used to define the critical lines that bound the lock-in state in the ($f_f,A$) space. For a given frequency, the critical amplitude $A=A_c$ can be determined by inspecting the evolution of the ratio  $a_{max}/x_{max}$. in Fig.~\ref{global_modes}(b, inset). As $f_f$ is fixed, the effective control parameter changes with $A$, so we expect  that $a_{max}\sim (A-A_c)^ {1/2}$ and $x_{max}\sim (A-A_c)^ {-1/2}$ which means that $a_{max}/x_{max}$ should behave linearly with $(A-A_c)$. We observe that the scaling holds near the critical value, while far from the critical lines it is modified  by higher non-linearities. We scale $a_{max}/x_{max}$  and the forcing amplitude $A$ to the distance to threshold with the critical value for each forcing frequency $(A-A_c)/A_c$. Fig.~\ref{global_modes}(b) resumes the three cases and a single line represents the linear behavior as $a_{max}/x_{max}\rightarrow 0$ for the control parameter $(A-A_c)/A_c$ as it attains the threshold.

\section{Drag estimation}
\label{drag_estim}
Previous works on this subject have addressed the problem of estimating forces from velocity fields only, employing different methods to include the contribution of the pressure field in the momentum balance in a control volume equation (a framework originally developed by \cite{wu1981}):
\begin{equation}\label{momentum1}
F=-\rho\frac{D}{Dt}\int_V \vec u dv + \int_S (-p\mathbf{I}+\mathbf{T})\cdot \vec n ds
\end{equation}

\noindent where $V$ is a control volume, $S$ its boundary, $\rho$ is the fluid density, $p$ is the pressure field, $\mathbf{I}$  the unit tensor and $\mathbf T=\mu(\nabla\vec u+\nabla^T\vec u)$ is the viscous stress tensor.  The pressure field can be obtained either by means of the Poisson equation (see e.g. \cite{fujisawa2005}), or integrating the Navier-Stokes (NS) equation along the control surface \cite{unal1997}. Considering for the pressure $p$ along a $s-$curve, $p(s)=p(s-ds)+\nabla p \cdot \vec ds$, the latter idea was further refined by \cite{kurtulus2007} who proposed using the NS equation only in the wake region, while adopting the Bernoulli equation in the surrounding slowly-evolving potential flow region. Thus,  $p(s)=\frac{\partial\phi}{\partial t}+p_0-\frac{1}{2}\rho|\vec u|^2$ which reduces the numerical error introduced by derivations.

\begin{figure}\centering
\includegraphics[width=0.8\linewidth]{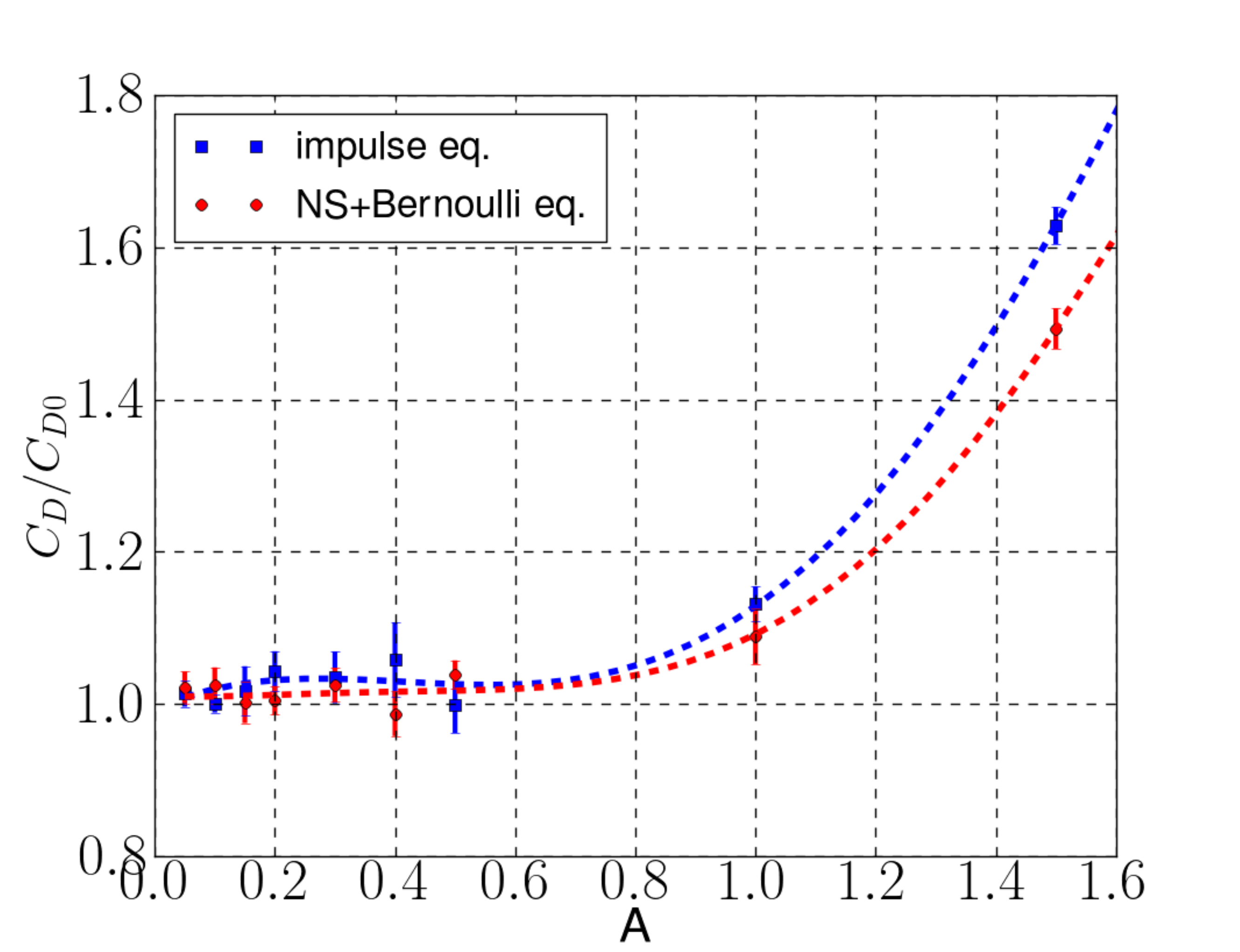}
\includegraphics[width=0.8\linewidth]{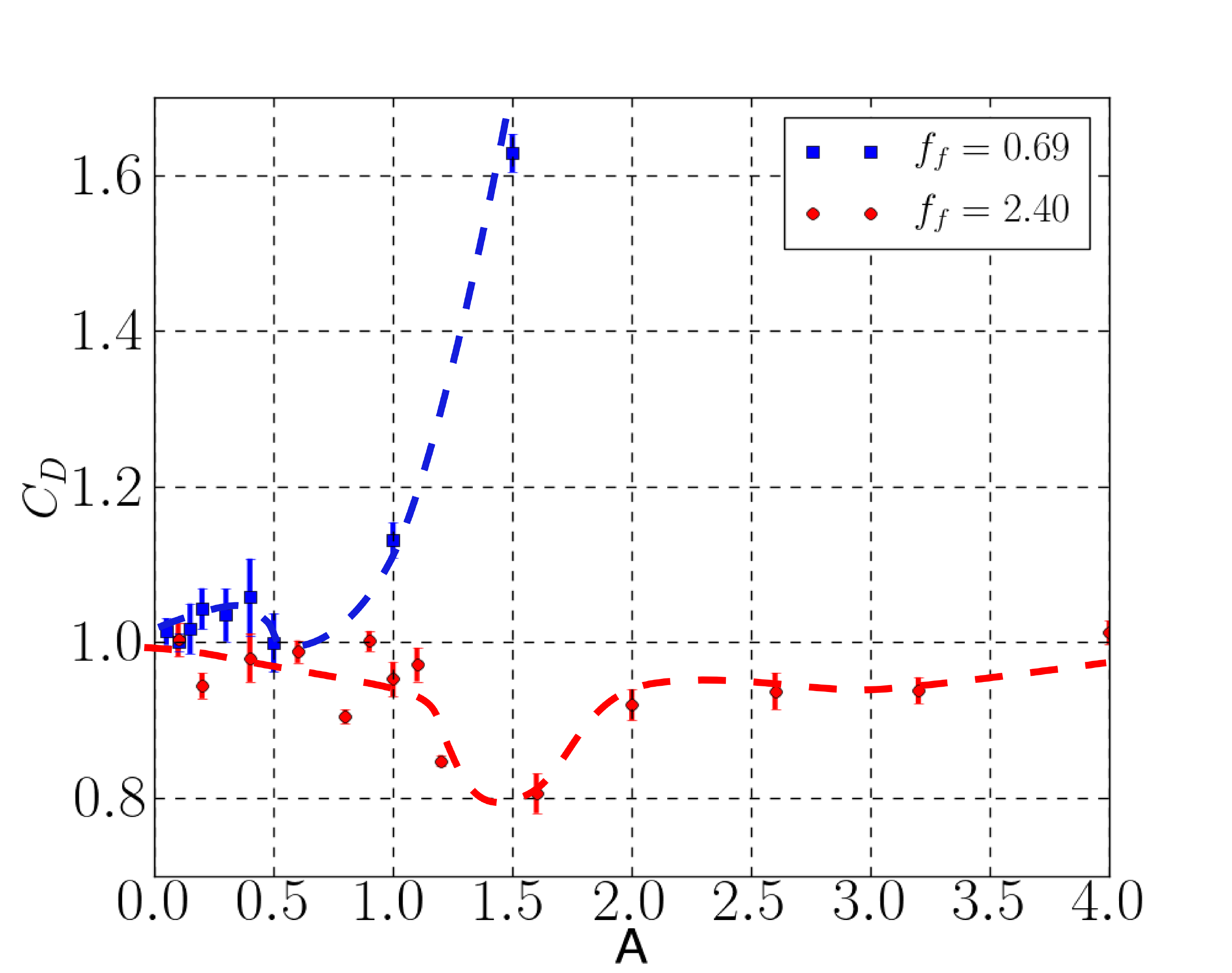}
\caption{(Color) Top: Drag coefficient for $f_f=0.69$ for Bernoulli-NS evaluation of pressure and  for the 'flux' Equation (\ref{noca}). $C_{D0}$ corresponds to the non-forced flow and fit curves show the general behavior for this case. Bottom: Drag coefficient for $f_f=0.69$ and for $f_f=2.40$.}
\label{drag_fig}
\end{figure}

Another approach has also been used to evaluate the force using only velocity fields and their derivatives (see e.g. \cite{noca1999,tan2005} ). It makes use of the the identities (see also \cite{saffman1992}):
\begin{equation}
 \frac{1}{\mathcal{N}-1}\int_V \vec x \times \vec \omega = \int_V \vec u dv +  \frac{1}{\mathcal{N}-1}\oint_S \vec x \times(\vec n \times\vec u)dS
\end{equation}
\begin{equation}
 \frac{D}{Dt}\oint_S\vec n \cdot\Phi dS=\oint_S\vec n\cdot\left[\frac{\partial \Phi}{\partial t}+\vec u_s(\nabla\cdot\Phi)\right]dS
\end{equation}

\noindent where $\Phi=[(\vec x \vec u )I - \vec x \vec u]$, so that Eq.~(\ref{momentum1}) leads to expressions where $p$ does not appear explicitly. For a 2D problem, the mean flow drag forces becomes:

\begin{eqnarray}\label{noca}
\left<F\right >&=&\oint_S \vec n\cdot {\big\{}  \frac{1}{2}\left<u^2\right>\mathbf{I} - \left<\vec u\vec u\right>-\left<\vec u(\vec x\times \vec \omega)\right> \\& & 
+\left[(\vec x\cdot\nabla\cdot\left<\mathbf{T}\right>)\mathbf{I}-\vec x\nabla\cdot\left<\mathbf{T}\right>\right]+\left<\mathbf{T}\right>   {\big\}}  dS \nonumber
\end{eqnarray}

\noindent where brackets indicate a time averaging procedure. We tested both the mixed Bernoulli-NS scheme, and the expression from Eq.~(\ref{noca}) using the present experimental velocity field measurements. The drag coefficient $C_D=  2 F_x / \rho  U_0^2 D$ shown on Fig.~\ref{drag_fig} (top)  was determined for the volume around the cylinder limited by $x=-2D$,  $x=3D$ and $y=\pm 4D$.  For the non-forced case, the reference value $C_{D0}=1.53$ is in  reasonable agreement with the literature (confinement effects on the $U_0$ value have been taken into account in the calculation of $C_{D0}$). The error was estimated from the rms value of $C_D$ for a variation of $1D$ on the control volume boundaries. The error is slightly lower for the impulse equation, so we adopted it to estimate the drag on the cylinder for different forcing parameters.

Fig.~\ref{drag_fig} (bottom) shows the evolution of the drag force as we increase the forcing amplitude $A$ for two  cases $f_f=0.69 f_0$ and $f_f=2.40 f_0$, representative of forcing below and above the natural frequency, respectively. For frequencies $f_f<f_0$ the drag force increases strongly after the lock-in threshold. Qualitatively different, for $f_>f_0$ the force decreases only when we are sufficiently near the threshold, attaining its minimum not far passed from $A_c$.  The maximum drag reduction of around 20\% is consistent with what has been reported in other experimental works \cite{tokumaru1991,thiria2006} as well as numerical simulations \cite{protas2002,bergmann2005}. Comparing the drag chart in the $(f_f,A)$ space obtained by \cite{bergmann2005} from numerical simulations with  Fig.~\ref{totalcasos} suggests that for $f_f>f_0$ the drag minimizes for regions close to the lock-in threshold, where the global fluctuations are reduced. On the other hand, maximum drag is obtained well inside the lock-in region for $f_f<f_0$ as in \cite{bergmann2005} (see Fig.~\ref{totalcasos}).

\section{Spectral analysis of the forced wake}
\label{spec_study}
\begin{figure}
\centering
\begin{overpic}[width=0.85\linewidth,trim= 14mm 9mm 5mm 0mm,clip,unit=1mm]%
{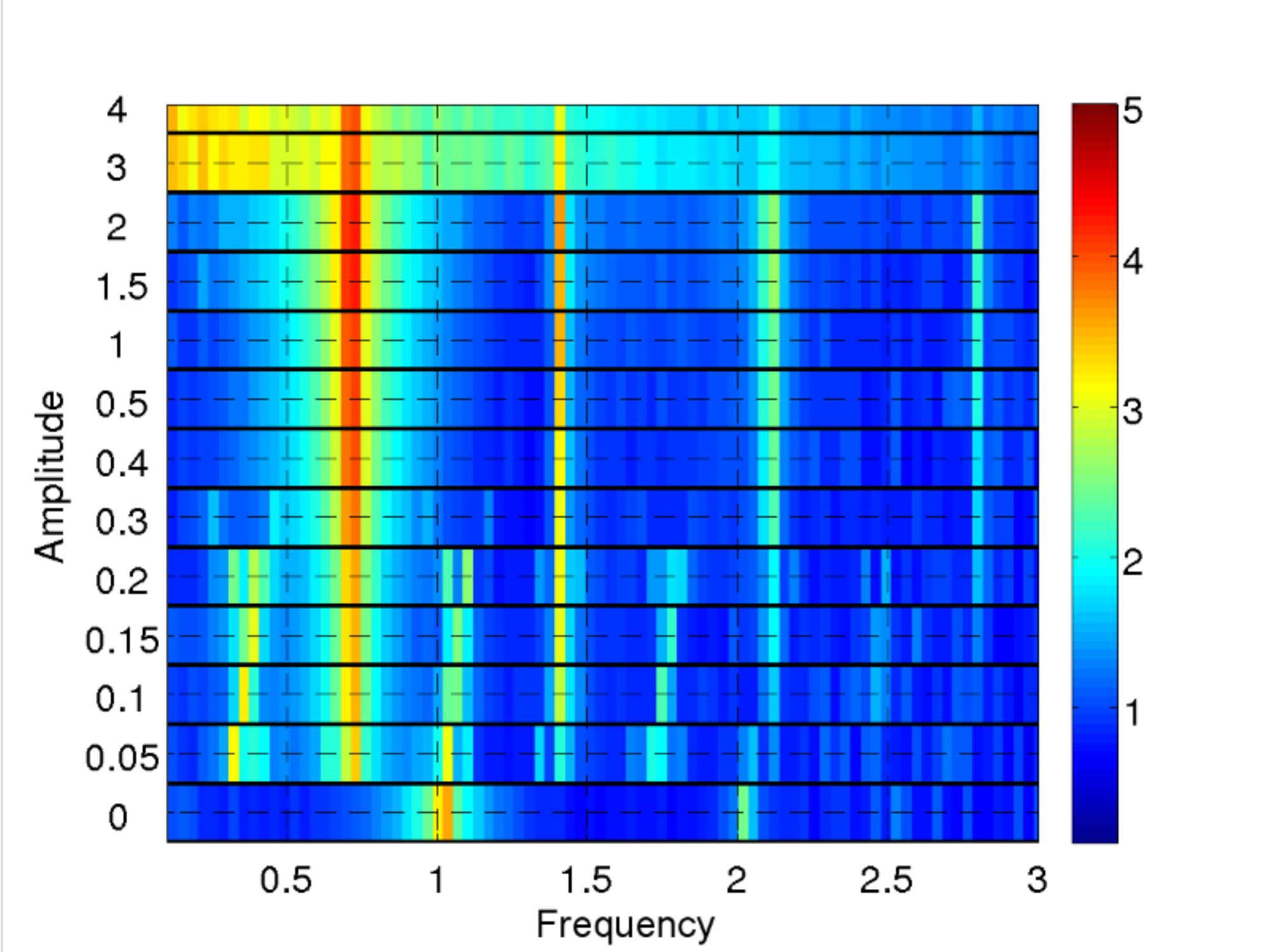}
\put(-4,35){\small A}
\put(45,-3){\small $f_f$}
\end{overpic}
\begin{overpic}[width=0.85\linewidth,trim= 14mm 9mm 5mm 0mm,clip,unit=1mm]%
{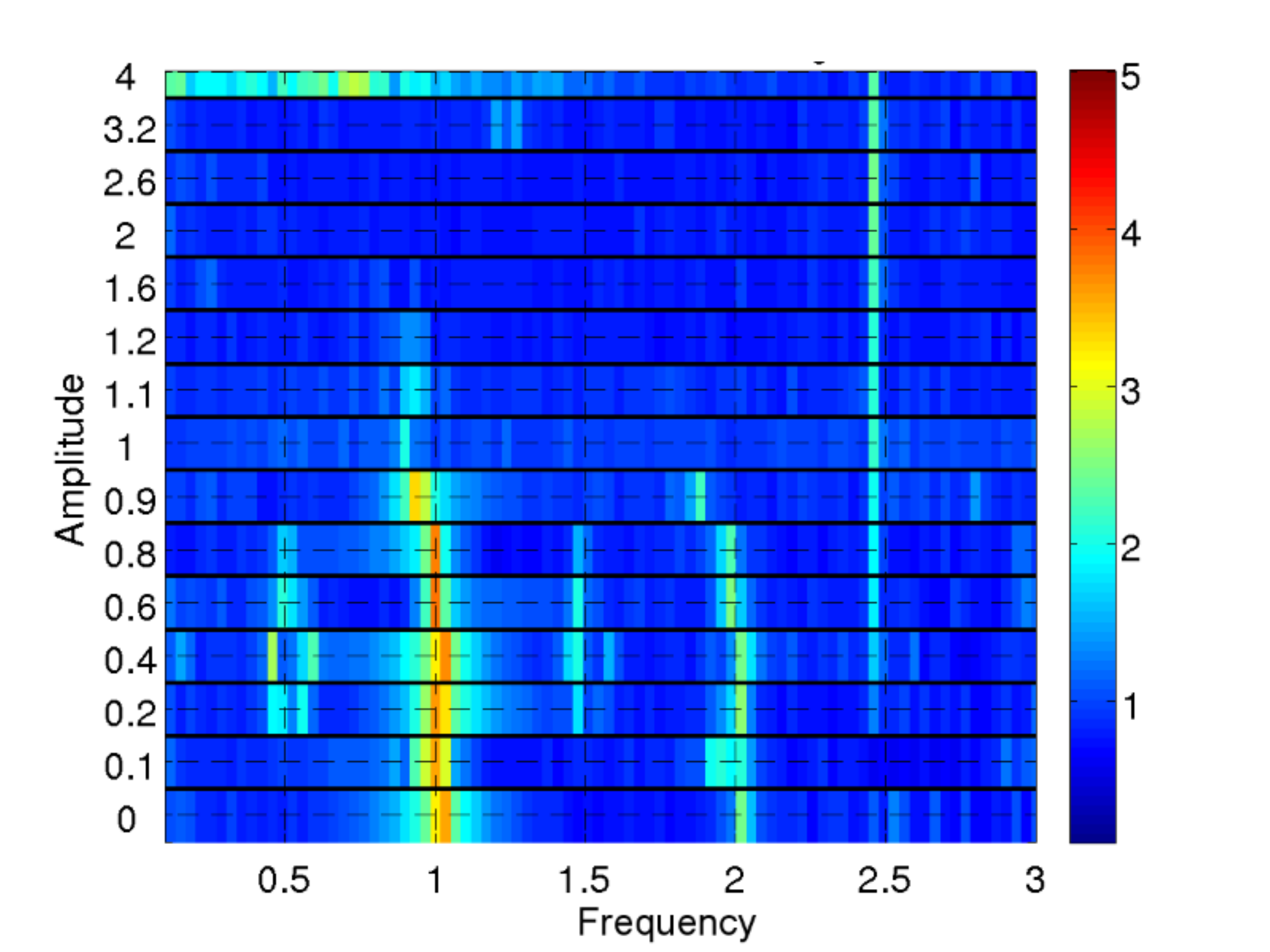}
\put(-4,35){\small A}
\put(45,-3){\small $f_f$}
\end{overpic}
\caption{(Color) Spectrograms for (Top) $f_f=0.69f_0$ and (Bottom) $f_f=2,40f_0$. The power density spectra are represented as colour levels (in a log scale) on a frequency content, forcing amplitude (nonlinear scale) map.}
\label{spectrogram}
\end{figure}

\begin{figure}
  \centering
 \subfigure[$A=0.10$]{\label{especf069-a}
\includegraphics[width=0.98\linewidth,trim= 15mm 2mm 5mm 10mm,clip]{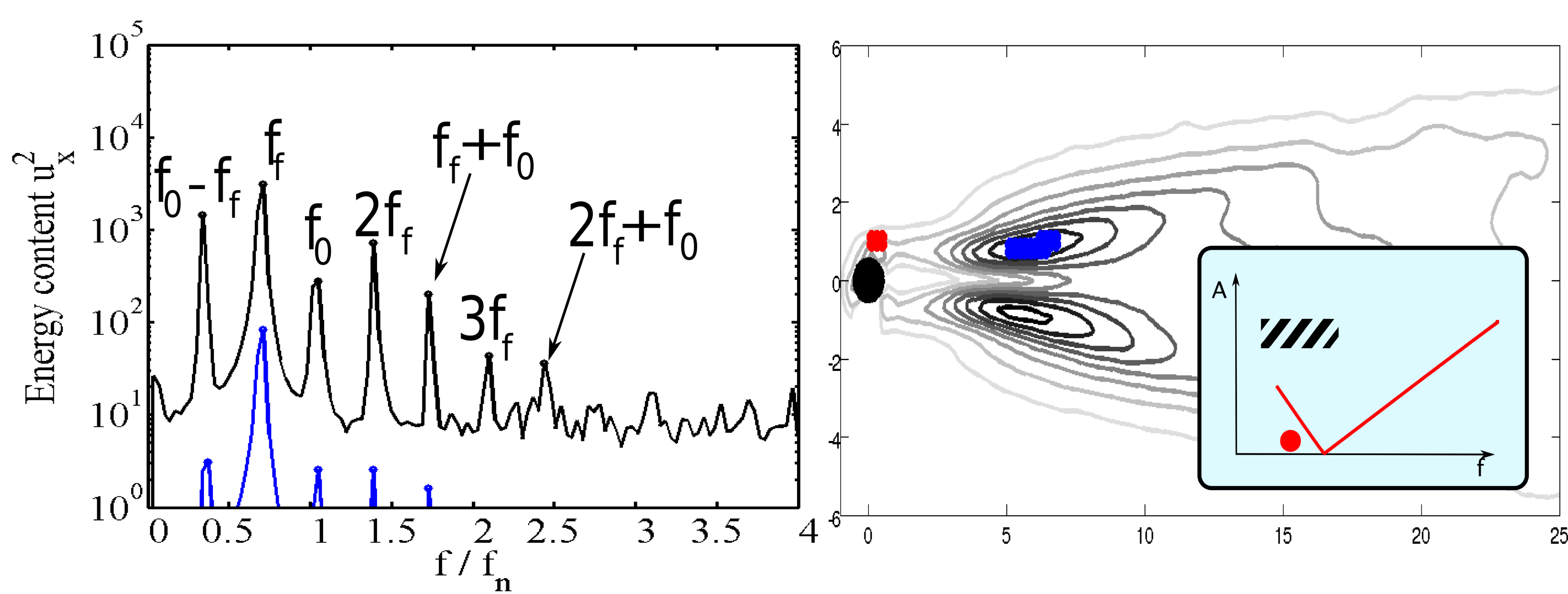}}\\
\subfigure[$A=0.40$]{\label{especf069-b}
\includegraphics[width=0.98\linewidth,trim= 15mm 2mm 5mm 10mm,clip]{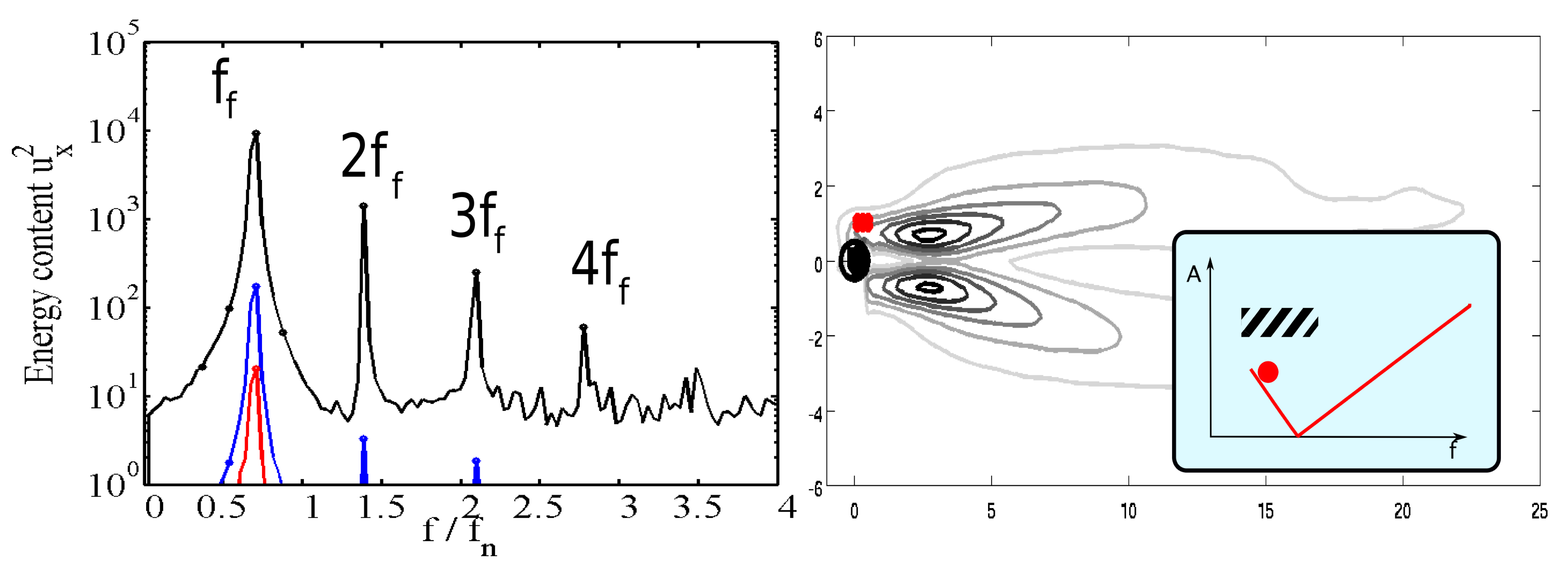}}\\
\subfigure[$A=3.00$]{\label{especf069-c}
\includegraphics[width=0.98\linewidth,trim= 15mm 2mm 5mm 10mm,clip]{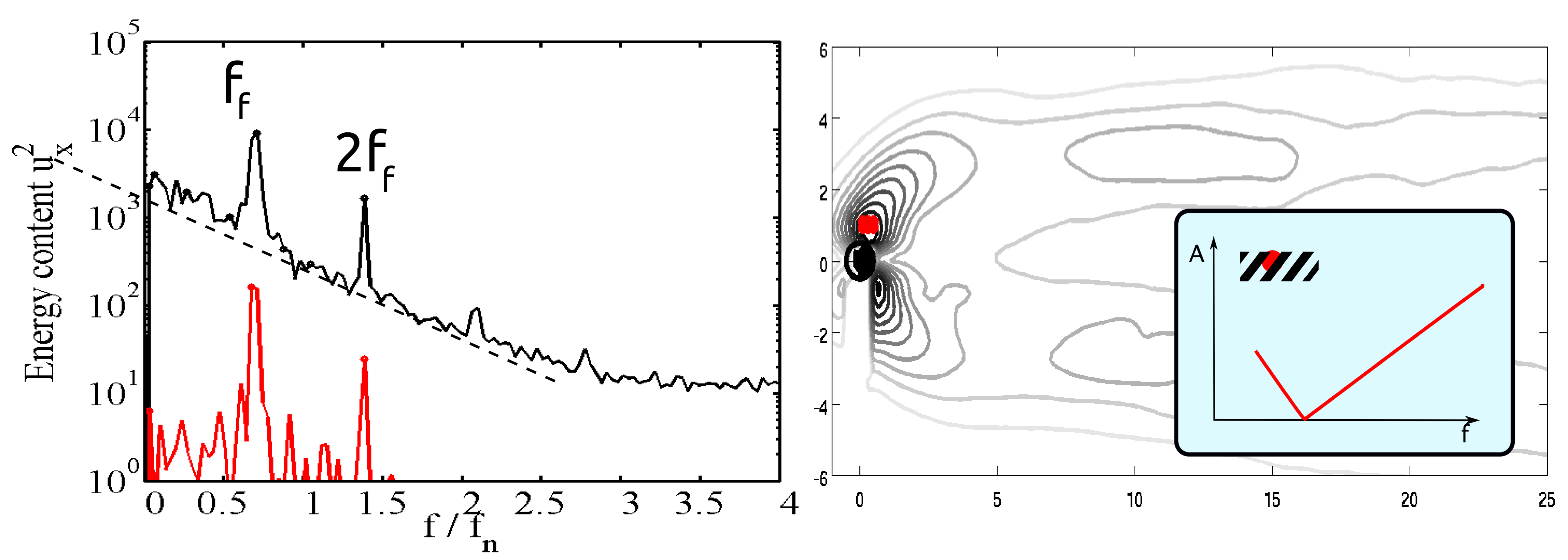}}
\caption{(Color) \textit{Left}: Power spectrum for a forcing frequency of $0.69f_0$ for three forcing amplitudes. Black Line: Global spectrum; Blue line: spectrum around the global mode maximum; Red: Spectrum in the cylinder vicinity. Right: Contour levels of  $rms(u_x)$ with the positions of the spectral probes marked in blue and red. The parameters of the forcing for each case are indicated in the lock-in diagram.}
\label{especf069}
\end{figure}

\begin{figure}
  \centering\vspace{.1cm}
 \subfigure[$A=0.40$]{\label{especf240-a}
\includegraphics[width=0.98\linewidth,trim= 15mm 0mm 5mm 7mm,clip]{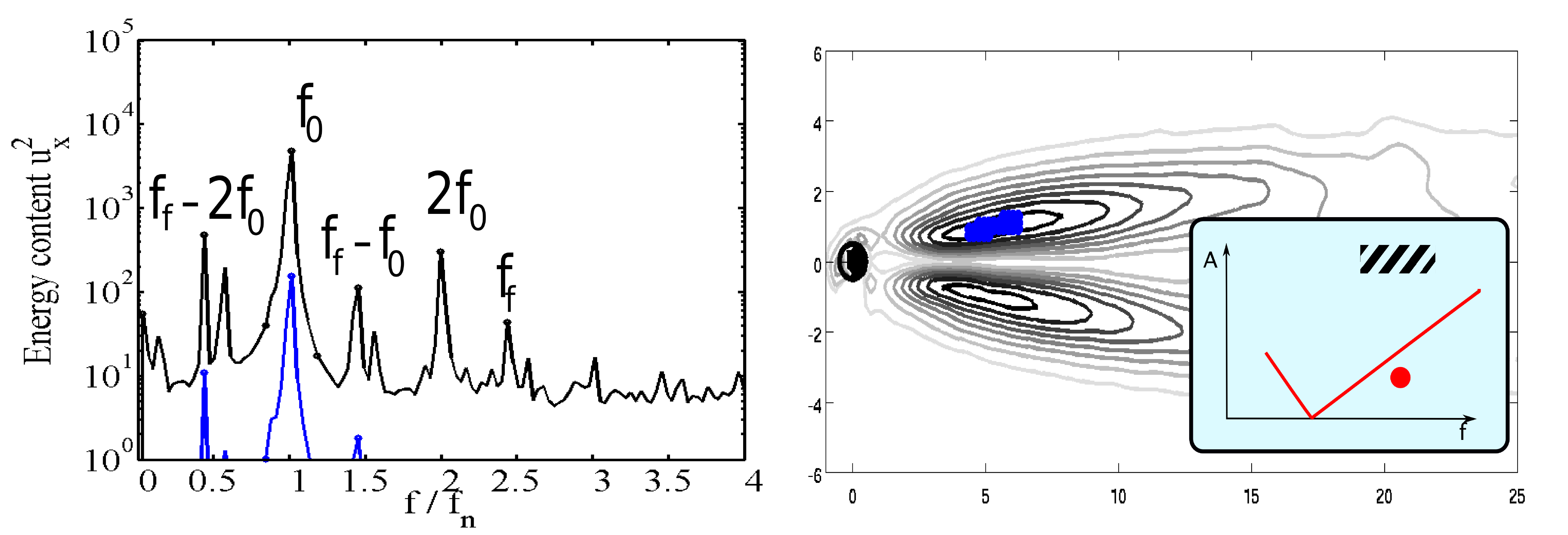}}\\\vspace{.2cm}
\subfigure[$A=1.60$]{\label{especf240-b}
\includegraphics[width=0.98\linewidth,trim= 15mm 0mm 5mm 5mm,clip]{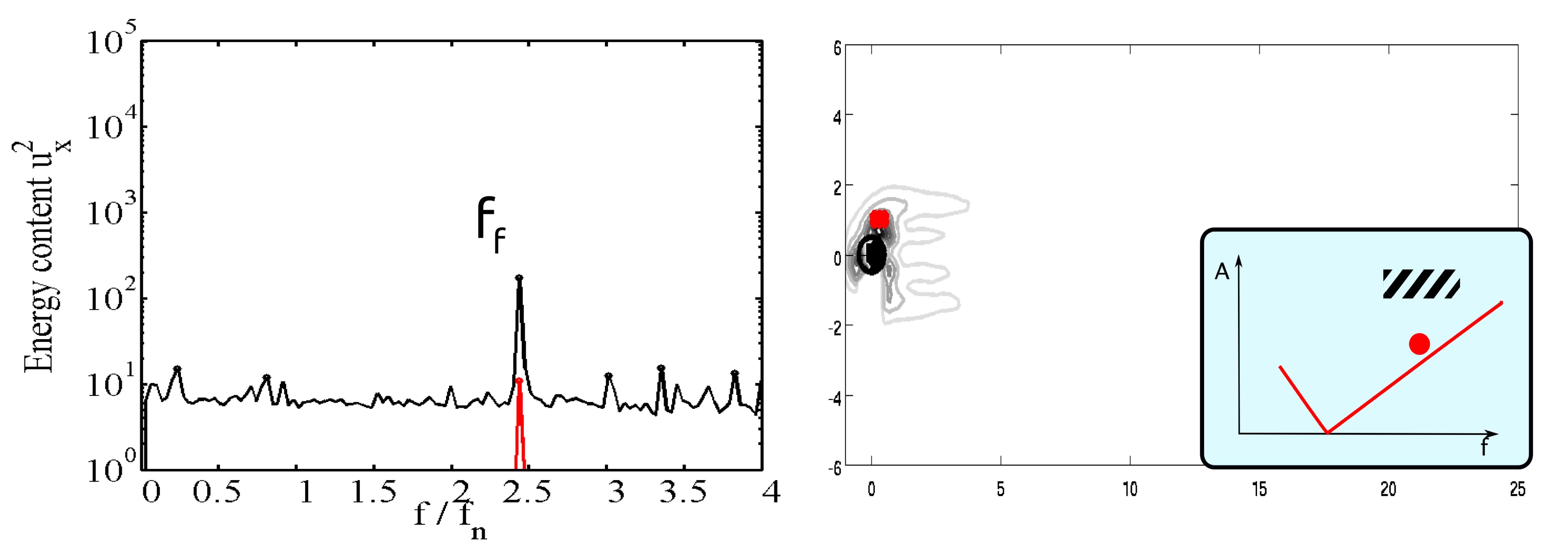}}\\
\subfigure[$A=4.00$]{\label{especf240-c}
\includegraphics[width=0.98\linewidth,trim= 15mm 0mm 5mm 7mm,clip]{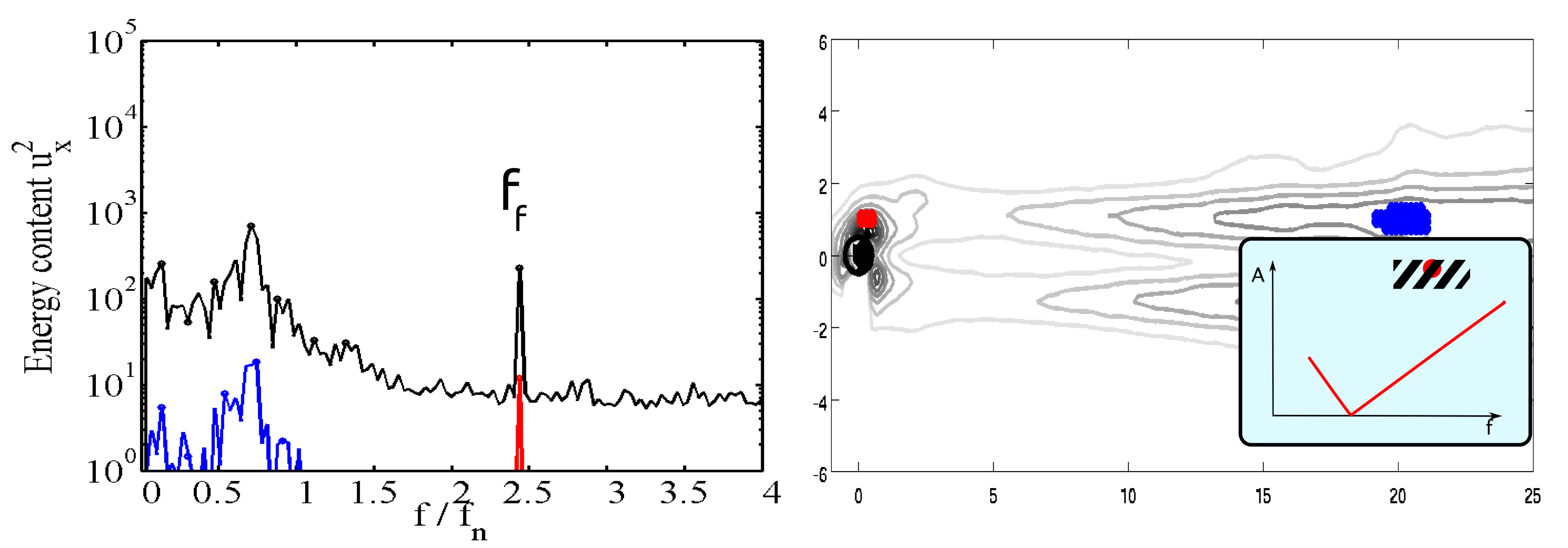}}
\caption{(Color) \textit{Left}: Power spectrum for a forcing frequency of $2.40f_0$ for three forcing amplitudes. Black Line: Global spectrum; Blue line: spectrum around the global mode maximum; Red: Spectrum in the cylinder vicinity. Right: Contour levels of  $rms(u_x)$ with the positions of the spectral probes marked in blue and red. The parameters of the forcing for each case are indicated in the lock-in diagram. }

\label{especf240}
\end{figure}

The lock-in threshold can be explored by inspecting the power spectral density on the flow domain. For each point in the parameter space we calculate the spectrum of $u_x$ for every point in the domain. The global spectrum is then obtained as the sum of all points spectra. In addition to this global spectrum that contains all frequencies present in the flow, we define two ``spectral probes'' in two regions chosen considering the spatial development of the flow: one in the cylinder vicinity, and the second one in the location of the maximum strength of velocity fluctuations.  A synthesis of all the global spectra is represented in the spectrograms shown in Fig.~\ref{spectrogram} for the same two forcing frequencies that have been analyzed in the previous sections. The spatial modification of the wake structure caused by the forcing is presented alongside the corresponding spectra in Figs.~\ref{especf069} and \ref{especf240}. For each forcing amplitude the root mean square of $u_x$ is displayed by its contour levels.

Fig.~\ref{spectrogram}(a) corresponds to a forcing frequency of $0.69f_0$ and represents the characteristic behaviour for $f_f<f_0$. The first rank ($A=0$) is the non-forced case where only the peaks on the natural frequency and its harmonics are observed. Next, under forcing, even for amplitudes a low as $A=0.05$, the spectrum is modified by the appearance of the forcing frequency, its harmonics and linear combinations of forcing and natural frequencies, $k_1 f_f+ k_2 f_0$ (where $k_1$ and $k_2$ are integers). Fig.~\ref{especf069-a} shows a typical power spectral density for this regime.  We observe also in Fig.~\ref{spectrogram}(a) that the natural frequency value increases when the forcing amplitude approaches the lock-in threshold.  The peak for $f_0$ decreases its intensity as the forcing amplitude increases, until the forcing frequency peak and its harmonics are the only present on the flow ($A>0.3$ for this case). As the global mode vanishes, the flow is locked on the forcing frequency (see Fig.~\ref{especf069-b}).
The peak on the forcing frequency continues to grow and its amplitude reaches about 10 times the non forced case.
For higher amplitudes ($A>2$ for this case) the spectrum becomes continuous (see Fig.~\ref{especf069-c}), with the forcing peaks distributed over a slope that clearly links higher energetic scales (low frequencies) to smaller scales (high frequencies). A $-2$ slope fits the curve, as it occurs in flows under strong rotational forcing \cite{morize2005}, unlike isotropic turbulence that fits with a $-5/3$ slope. This state is attained at high amplitudes for all forcing frequencies $f_f<f_0$ for high amplitudes and is, to our knowledge, a novel observation in the context of open flows.
The spectrogram in Fig.~\ref{spectrogram}(b) represents the spectrum variation for a forcing frequency $f_f=2.40 f_0$. Again, for amplitudes lower than those near the lock-in threshold, the flow presents a spectrum with discrete peaks $k_1 f_f+ f_2 f_0$ (Fig.~\ref{especf240-a}). The natural frequency peak decreases its energy as the forcing amplitude increases. When approaching the threshold ($A>1.1$ for this case) the natural frequency diminishes its value to $0.88f_0$. This has been discussed by \cite{thiria2009} as a consequence of the wake re-stabilization that changes the linear global frequency selected by the base flow. Above the lock-in threshold, like on Fig.~\ref{especf240-b} for $A=1.60>A_c$, the flow contains only the forcing frequency and its harmonics, and energy level of the fluctuations is much lower, about 10 times smaller than the corresponding for the non forced case. Another threshold appears as for $A>3.2$ where the spectrum becomes continuous (Fig.~\ref{especf240-c}), but this is owing to strong fluctuations that come from the far wake, unlike the precedent case where they were generated directly by the forcing so the phenomenon is qualitatively different. We notice from the spectrogram of Fig.~\ref{especf240-b}  that a peak on $\sim f_f/2$ first emerges prior to the transition to a continuous spectrum for higher amplitudes.\\
It is worth mentioning  that similar dynamics  have been reported in the case of enclosed swirling flows under harmonic forcing \cite{lopez2008} as well as on other forced systems \cite{chiffaudel1987}. Indeed, the quasiperiodic behavior for $f_f$ and $f_0$  that characterizes the transition to lock-in has been reported by \cite{lopez2008}, who have described in detail the period doubling process.  Further investigations on $Re<Re_c$ may shed more light on the mechanisms of the transition process and the similarities  of enclosed swirling flow studied by \cite{cui2009}.

\section{Conclusions}
Spectral analysis was shown to be a useful tool for the analysis of PIV experimental data from a forced cylinder wake. Lower Reynolds number than previous works, allow us to characterize the flow's critical behavior more accurately.
 In addition to refining previous studies on the global mode properties of forced wakes and revisiting their scaling properties, the present results have in particular allowed to shed new light on previously unexplored phenomena related to the transitions between globally unstable and locked-in states in the parameter space of the forcing $(f,A)$. The appearance of a continuous spectrum for large forcing amplitudes was observed in two different situations with presumably two different physical explanations: for $f<f_0$, the large forcing amplitude determines each vortex shed in the near wake to be very intense and thus to be destabilized, split and mix rapidly giving rise to a turbulence-like window in the parameter space. This regime is very easy to obtain and can be used extensively to generate turbulent behavior at moderate Reynolds numbers. On the other hand, for  $f>f_0$, passed the lock-in threshold, under increasing forcing amplitude, the flow is subject to large fluctuations coming from the far wake.\\
 Finally, we have evaluated the drag force from the velocity field bringing experimental evidence to suggest a relationship between the drag minimum and the lock-in threshold.
\vspace{-10mm}
\acknowledgments
We acknowledge gratefully B. Thiria for useful discussions. The present work was supported by the Franco-Argentinian Associated Laboratory in the Physics and Mechanics of Fluids (LIA PMF-FMF).


\end{document}